\documentclass[aps,prd,preprint,eqsecnum,tightenlines,floats,
groupedaddress,showpacs]{revtex4}
\usepackage{amssymb} 
\usepackage{hyperref} 
\usepackage{graphicx}
\usepackage{subfigure}

\begin{document}


\date{January 16, 2018}

\title{Gravitational self-interactions of a 
degenerate \\ quantum scalar field}

\author{Sankha S. Chakrabarty, Seishi Enomoto, Yaqi Han, 
Pierre Sikivie and Elisa M. Todarello}
\affiliation{Department of Physics, University of Florida, 
Gainesville, FL 32611, USA}

\begin{abstract}

We develop a formalism to help calculate in quantum field 
theory the departures from the description of a system 
by classical field equations.  We apply the formalism to 
a homogeneous condensate with attractive contact interactions 
and to a homogeneous self-gravitating condensate in critical 
expansion.  In their classical descriptions, such condensates 
persist forever.  We show that in their quantum description, 
parametric resonance causes quanta to jump in pairs out of 
the condensate into all modes with wavevector less than some 
critical value.  We calculate in each case the time scale 
over which the homogeneous condensate is depleted, and after 
which a classical description is invalid. We argue that the 
duration of classicality of inhomogeneous condensates is 
shorter than that of homogeneous condensates.

\end{abstract}

\maketitle

\section{Introduction}

The identity of dark matter remains one of the foremost questions in 
science today \cite{PDM}.  One of the leading candidates is the QCD 
axion, which has the double virtue of solving the strong CP problem 
of the standard model of elementary particles \cite{axion,invax} and of 
being naturally produced with very low velocity dispersion in the early 
universe \cite{adm}, so that it behaves as cold dark matter from the 
point of view of structure formation \cite{ipser}.  Several other  
candidates, called axion-like particles (ALPs) or weakly interacting 
slim particles (WISPs), have properties similar to axions as far as the 
dark matter problem is concerned \cite{Arias}.  ALPs with mass of order 
$10^{-21}$ eV, called ultra-light ALPs (ULALPs), have been proposed as
a solution to the problems that ordinary cold dark matter is thought to 
have on small scales \cite{ULALP}.  Axion dark matter has enormous 
quantum degeneracy, of order $10^{61}$~\cite{CABEC} or more. The 
degeneracy of ULALP dark matter is even higher \cite{Christ}.  In 
most discussions of axion or ALP dark matter, the particles are 
described by classical field equations.  The underlying assumption 
appears to be that huge degeneracy ensures the correctness of a 
classical field description.

However it was found in refs. \cite{CABEC,axtherm,Saik,Berges} that 
cold dark matter axions thermalize, as a result of their gravitational 
self-interactions, on time scales shorter than the age of the universe
after the photon temperature has dropped to approximately one keV.  
When they thermalize, all the conditions for their Bose-Einstein 
condensation are satisfied and it is natural to assume that this is
indeed what happens.  Axion thermalization implies that the axion 
fluid does not obey classical field equations since the outcome
of thermalization in classical field theory is a UV catastrophe, 
wherein each mode has average energy $k_B T$ no matter how high the 
mode's oscillation frequency, whereas the outcome of thermalization 
of a Bosonic quantum field is to produce a Bose-Einstein distribution.
On sufficiently short time scales, the axion fluid does obey classical 
fields equations.  It behaves then like ordinary cold dark matter on all 
length scales longer than a certain Jeans length \cite{Khlopov,Bianchi}; 
see Eq.~(\ref{Jeans}) below.  However, on longer time scales, the axion 
fluid thermalizes.  When thermalizing, the axion fluid behaves differently 
from ordinary cold dark matter since it forms a Bose-Einstein condensate, 
i.e. almost all axions go to the lowest energy state available to them.   
Ordinary cold dark matter particles, weakly interacting massive particles 
(WIMPs) and sterile neutrinos \cite{PDM} do not have that property.

Axion thermalization has implications for observation.  It was found 
\cite{axtherm} that the axions which are about to fall into a galactic 
potential  well thermalize sufficiently fast that they almost all go 
to their lowest energy state consistent with the total angular momentum 
they acquired from tidal torquing.  That state is one of rigid rotation 
in the angular variables (different from rigid body rotation but similar 
to the rotation of water going down a drain), implying that the velocity 
field has vorticity ($\vec{\nabla}\times\vec{v} \neq 0$).  In contrast, 
ordinary cold dark matter falls into gravitational potential wells with 
an irrotational velocity field \cite{inner}.  The inner caustics of 
galactic halos are different in the two cases.  If the particles fall 
in with net overall rotation the inner caustics are rings whose 
cross-section is a section of the elliptic umbilic catastrophe, 
called caustic rings for short \cite{crdm,sing}.  If the particles 
fall in with an irrotational velocity field, the inner caustics have 
a tent-like structure \cite{inner} quite distinct from caustic rings.  
Observational evidence had been found for caustic rings.  The evidence
is summarized in ref. \cite{MWhalo}. It was shown \cite{case,Banik} that 
axion thermalization and Bose-Einstein condensation explains the evidence 
for caustic rings of dark matter in disk galaxies in detail and in all its
aspects, i.e. it explains not only why the inner caustics are rings and why 
they are in the galactic plane but it also correctly accounts for the overall 
size of the rings and the relative sizes of the several rings in a single halo.  
Finally it was shown that axion dark matter thermalization and Bose-Einstein 
condensation provide a solution \cite{Banik} to the galactic angular momentum 
problem \cite{Burkert}, the tendency of galactic halos built of ordinary 
cold dark matter (CDM) and baryons to be too concentrated at their centers.  
An argument exists therefore that the dark matter is axions, at least in 
part.  Ref. \cite{Banik} estimates that the axion fraction of dark matter 
is 35\% or more.

All the above claimed successes notwithstanding, axion thermalization 
and Bose-Einstein condensation is a difficult topic from a theoretical 
point of  view.  Thermalization by gravity is unusual because gravity 
is long-range and, more disturbingly, because it causes instability.  
Bose-Einstein condensation means that a macroscopically large number 
of particles go to their lowest energy state.  But if the system is 
unstable it is not clear in general what is the lowest energy state.  
The idea that dark matter axions form a Bose-Einstein condensate was 
critiqued in refs. \cite{Davidson,Davidson2,Guth}.  It was concluded 
in ref. \cite{Guth} that ``while a Bose-Einstein condensate is formed, 
the claim of long-range correlation is unjustified."

In Section II of this paper, we aim to clarify aspects of Bose-Einstein
condensation that appear to cause confusion, at least as far as dark 
matter axions are concerned.  One issue is whether a Bose-Einstein 
condensate needs to be homogeneous (i.e. translationally invariant as 
is a condensate of zero momentum particles).  We answer this question 
negatively.  A Bose-Einstein condensate can be, and generally is, 
inhomogeneous. Nonetheless, merely by virtue of being a Bose-Einstein 
condensate, it is correlated over its whole extent, and its extent 
can be arbitrarily large compared to its scale of inhomogeneity.  

A second question is whether Bose-Einstein condensation can be described 
by classical field equations.  We state the following to be true.  The 
behavior of the condensate is described by classical field equations on 
time scales short compared to its rethermalization time scale.  However 
when the condensate rethermalizes, as it must if situated in a time-dependent 
background or if it is unstable, it does not obey classical field equations.  
A phenomenon akin to Bose-Einstein condensation does exist in classical 
field theory when a UV cutoff is imposed on the wave-vectors, i.e. all 
modes with wavevector $k > k_{\rm max}$ are removed from the theory.  
$k_{\rm max}$ is related to the critical temperature $T_{\rm crit}$ 
for Bose-Einstein condensation in the quantum field theory.  We emphasize 
however that the relationship $k_{\rm max}$ and $T_{\rm crit}$ necessarily 
involves a constant, such as $\hbar$, with dimension of action.  Furthermore, 
if we replace the quantum axion field by a cutoff classical field, even if 
a phenomenon similar to Bose-Einstein condensation does occur, there is no 
proof or expectation that the cutoff classical theory reproduces the other 
predictions of the quantum theory.  In particular, the phenomenology of 
caustic rings cannot be reproduced in the classical field theory, with or 
without cutoff, because vorticity (the circulation of the velocity field 
along a closed curve) is conserved in classical field theory.   In contrast, 
the production of vorticity and the appearance of caustic rings is the 
expected behaviour of the quantum axion fluid.

A broadly relevant question is the following: over what time scale is a 
classical description of a highly degenerate but self-interacting Bosonic 
system valid?  We call that time scale the duration of classicality of 
the system.  Two of us (PS and ET) have recently \cite{simtherm} addressed 
this by numerically integrating the equations of motion of a toy model 
consisting of five interacting quantum oscillators.  In all the cases 
simulated, the duration of classicality was found to be shorter or at 
most of order the thermal relaxation time $\tau$.  A summary of the 
results of ref. \cite{simtherm} is included in subsection II.2.  We 
add however new simulations in which only one of the five oscillators 
is excited in the initial state.  According to its classical evolution, 
this state persists forever.  According to its quantum evolution, the 
state has a finite lifetime because the quanta jump out of the initially 
excited oscillator into the four others.  The new simulations parallel our 
analytical treatment of the homogeneous condensate with attractive contact 
interactions in Section III and of the homogeneous self-gravitating 
condensate in critical expansion in Section IV.  

In Section III we develop a formalism to help calculate the quantum 
evolution of a scalar field that is described in its initial state 
by a classical solution.  A classical solution corresponds to one 
mode of the quantum field.  In the initial state all quanta are 
placed into that single mode.  However the quantum field has an 
infinite number of other modes.  We expand the quantum field into 
a complete orthonormal set of modes built around an arbitrary  
classical solution.  We derive the Hamiltonian in terms of the 
associated creation and annihilation operators.  The formalism 
is applicable to any condensate described by a classical solution.  

We first apply the formalism to the homogeneous condensate in 
$\lambda \phi^4$ theory, with $\lambda > 0$ and $\lambda < 0$, 
in Section III.  The condensate is unstable when $\lambda < 0$.  
Nonetheless, in its classical description, the homogeneous 
condensate persists forever.  In its quantum description, the 
homogeneous condensate is depleted by parametric resonance.  
We obtain the time scale over which the condensate is depleted, 
which is in effect its duration of classicality.  In Section IV 
we apply the formalism to a homogeneous self-gravitating condensate 
in critical expansion.  Again, in its classical description, the 
condensate persists forever whereas it is depleted by parametric 
resonance in its quantum description.  Here too we derive the time 
scale over which the condensate is depleted, and after which a 
classical description is invalid.  In the self-gravity case, the 
instability grows as a power of time whereas it grows exponentially 
fast in $\lambda \phi^4$ theory with $\lambda < 0$.  The results we 
derive in Section III and IV are all exact statements in the limit 
where the number of quanta $N$ in the condensate is very large 
compared to the number of quanta not in the condensate.

Although we only analyze the behaviour of homogeneous condensates
in this paper, we expect our conclusions to apply to inhomogeneous 
condensates as well.  Indeed, a homogeneous condensate can be seen 
as a limiting case of inhomogeneous condensates.  Since homogeneous
condensates are depleted by parametric resonance, the same must be
true for inhomogeneous condensates, at least in the limit of small
deviations away from homogeneity.  In fact in our simulations of 
the five oscillator toy model we find that the condensates which 
persist forever according to their classical evolution are the 
condensates with the longest duration of classicality in their 
quantum evolution.  We explain this result on the basis of 
analytical arguments.  By analogy, we expect inhomogeneous 
condensates to have shorter durations of classicality than 
homogeneous ones.

Related topics were discussed in two recent papers \cite{Berges2,Dvali}.
Inter alia, ref. \cite{Berges2} solves the classical equations of motion 
for an initially almost homogeneous condensate with attractive contact
interactions numerically on a lattice.  If it were strictly homogeneous, 
the condensate would persist forever.  Perturbations are introduced to 
mimick quantum fluctuations.  As the perturbations grow, the condensate 
is depleted in a manner which is qualitatively consistent with our 
quantum field theory treatment.  Ref. \cite{Dvali} discusses, as we do, 
the duration of classicality of the cosmic axion fluid.  The conclusions 
of ref. \cite{Dvali} differ from ours.

A brief outline of our paper is as follows.  In Section II, we discuss
Bose-Einstein condensation and analyze four issues which may cause
confusion when the interactions are attractive, as is the case for 
dark matter axions.  In Section III, we introduce a formalism to 
calculate in quantum field theory the departures from a description 
of a system by classical field equations.  We apply it to the homogeneous 
condensate in $\lambda \phi^4$ theory in the repulsive ($\lambda > 0$) and 
attractive ($\lambda < 0$) cases.  In Section IV, we apply the formalism 
to a homogeneous self-gravitating condensate in critical expansion.  In 
Section V, we summarize our conclusions.

\section{Bose-Einstein condensation}

This section gives a brief description of the phenomenon of 
Bose-Einstein condensation, emphasizing the necessary and 
sufficient conditions for its occurrence, and the reason why 
it occurs.  We follow this with a discussion of four subtopics 
which appear occasionally to cause confusion in discussions of
Bose-Einstein condensation of dark matter axions.

Consider a system of $N$ identical bosons in thermal equilibrium 
under the constraint that the total number of particles is conserved.  
A standard textbook derivation yields the average occupation number 
$\langle {\cal N}_j \rangle$ of particle state $j$ in the limit of 
a huge number of particles (the so-called thermodynamic limit):
\begin{equation}
\langle {\cal N}_j \rangle =
{1 \over e^{{1 \over T}(\epsilon_j - \mu)} - 1}
\label{Bose}
\end{equation}
where $T$ is the temperature, $\mu$ the chemical
potential, and $\epsilon_j$ ($j = 0,1,2,3 ...$) the 
energy of particle state $j$.  We will assume 
that the particle states are ordered so that
$\epsilon_0 < \epsilon_1 < \epsilon_2 < ...~$.  The 
$\langle {\cal N}_j \rangle$ maximize the system entropy 
for a given total energy $E = \sum_j {\cal N}_j \epsilon_j$
and total number of particles $N = \sum_j {\cal N}_j$.  Since
all $\langle {\cal N}_j \rangle \geq 0$, it is necessary that 
$\mu < \epsilon_0$ for Eq.~(\ref{Bose}) to make sense.  On the 
other hand, the total number of particles 
$N(T,\mu) = \sum_j \langle {\cal N}_j \rangle$
is an increasing function of $\mu$ for fixed $T$ since each
$\langle {\cal N}_j \rangle$ is.  So, if $N$ is increased 
while $T$ is held fixed, $\mu$ must increase but it can not 
become larger than $\epsilon_0$.  In the systems of interest to 
us, the total number of particles in excited ($j>0$) states 
has for $\mu = \epsilon_0$ a finite value 
\begin{equation}
N_{\rm ex}(T,\mu=\epsilon_0) = 
\sum_{j>0} {1 \over e^{{1 \over T}(\epsilon_j - \epsilon_0)} - 1}~~\ .
\label{ttp}
\end{equation}
(In one and two spatial dimensions, $N_{\rm ex}(T,\mu=\epsilon_0)$
may be infinite because of an infrared divergence but this comment is 
not relevant to the systems in three spatial dimensions that interest 
us.) Consider what happens when, at fixed $T$, $N$ is made larger than 
$N_{\rm ex}(T,\mu=\epsilon_0)$.  The only possible system response is 
for the extra $N - N_{\rm ex}(T,\mu=\epsilon_0)$ particles to go to 
the ground state ($j=0$).  Indeed the average occupation number  
$\langle {\cal N}_0 \rangle$ of that state becomes arbitrarily 
large as $\mu$ approaches $\epsilon_0$ from below.

From the above we deduce four conditions for Bose-Einstein condensation:
i) the system comprises a large number of identical bosons, hereafter
called particles for short,  ii) the number of particles is conserved, 
iii) the particles are sufficiently degenerate, and iv) the particles are 
in thermal equilibrium.  The number of particles has to be sufficiently 
large (condition i) for the system to be in the thermodynamic limit.   
The number of particles has to be conserved (condition ii) but only on 
the time scale of thermalization.  It need not be absolutely conserved.  
For example, it is irrelevant to Bose-Einstein condensation in dilute gases 
whether baryons are absolutely stable.  The only thing that matters is 
that they are stable on the time scale of the condensation process.  
(Whether they decay tomorrow is irrelevant to their condensation this 
minute.)  Dark matter axions are not absolutely stable since they decay 
to two photons.  However their lifetime is much longer than the age of 
the universe and so is the timescale of all other axion number changing 
processes.  Condition iii) is satisfied if the degeneracy, i.e. the 
average occupation number  $\langle {\cal N}_j \rangle$ of those states 
that are occupied, is larger than some critical number of order one.
For systems in thermal equilibrium this condition is the requirement 
that the temperature is lower than some critical temperature.  The 
critical temperature is such that the interparticle distance is of 
order the thermal de Broglie wavelength. Thermal equilibrium is generally 
taken for granted in discussions of Bose-Einstein condensation in liquid 
$^4$He and dilute gases because these systems thermalize very quickly.  
For these systems, condition iv) is readily satisfied but condition iii) 
is difficult to achieve because of the low temperatures required.  The 
reverse situation pertains to dark matter axions.  Axions thermalize only 
very slowly, perhaps on the timescale of the age of the universe, or not 
at all, because axions are very weakly interacting.  On the other hand, 
their quantum degeneracy is enormous with ${\cal N} \sim 10^{61}$.  For 
these reasons we state condition iii) independently and ahead of 
condition iv).

Thermalization involves interactions and requires time.  We define the 
relaxation time $\tau$ to be the time scale over which the distribution 
$\{{\cal N}_j\}$ of the particles over the particle states changes 
completely, each ${\cal N}_j$ changing by order 100\%.  Whether condition 
iv) for Bose-Einstein condensation is satisfied is an issue of time scales.   
Let us assume that the first three conditions are satisfied and that the 
system is out of equilibrium, i.e. the system is in a state of entropy 
less than allowed.  The system will then thermalize on the time scale 
$\tau$, increasing its entropy.  It forms a BEC because the state of 
highest entropy, given that the first three conditions are satisfied,
is one in which a fraction of order one of the particles is in the 
lowest energy available state and the remaining particles in a thermal 
distribution in excited states.  The entropy increases when a BEC forms.  
The process is irreversible.

We now discuss four aspects of Bose-Einstein condensation that appear 
sometimes to be sources of confusion in the literature, and must be 
clarified especially in the context of cosmic axion Bose-Einstein 
condensation. 

\subsubsection{Quantum mechanics is essential}

It is possible within classical field theory to produce a phenomenon 
similar to Bose-Einstein condensation by introducing a cutoff $k_{\rm UV}$ 
on the wavevectors of the field modes.   Indeed the classical physics analog
of Eq.~(\ref{Bose}) is 
\begin{equation}
\langle {\cal N}_j \rangle = {T \over \epsilon_j - \mu}~~\ .
\label{class}
\end{equation}
When $\mu$ approaches $\epsilon_0$ from below, provided $T \neq 0$,
$\langle {\cal N}_0 \rangle$ diverges as it does for the Bose-Einstein 
distribution.  However in classical field theory the energy gets 
distributed equally over all field modes.  If there is no cutoff the 
specific heat per unit volume diverges because, in any finite volume, 
the field has an infinite number of modes with large wavevectors 
$\vec{k}$. In other words, $T = 0$ in any finite volume containing
finite energy in thermal equilibrium.

To produce a phenomenon resembling Bose-Einstein condensation in classical 
field theory, a cutoff is introduced by hand with value fixed so that 
\begin{equation} 
T_{\rm crit} \sim {k_{\rm UV}^2 \over 2m} 
\label{ghp} 
\end{equation} 
where $m$ is the boson mass and $T_{\rm crit}$ the critical 
temperature that is naturally present in the quantum theory.  
The number of modes per unit volume is finite then, of order 
$k_{\rm UV}^3/(2 \pi)^3$.  As a result $T \neq 0$ in the cutoff
field theory and ${\cal N}_0 \rightarrow \infty$ when $\mu$ 
approaches $\epsilon_0$ from below.  However, this does not 
mean that the cutoff classical field theory has any validity 
beyond producing some form of Bose-Einstein condensation.  The 
cutoff is not meant to be present in any real sense.  In general 
the cutoff classical field theory differs from the quantum field 
theory, and when the two make different predictions it is the 
latter that is to be believed not the former. In particular, as 
discussed in ref. \cite{Banik}, the classical theory conserves 
vorticity, i.e. the circulation of the velocity field along a 
closed path $\Gamma$ \begin{equation}
{\cal C}[\Gamma] \equiv \oint_\Gamma~d\vec{r}\cdot\vec{v}(\vec{r}, t)~~\ ,
\label{circ}
\end{equation}
whereas the quantum theory does not.  Conservation of vorticity in 
classical field theory follows from the continuity and single-valuedness 
of the wavefunction, and holds whether or not a wavevector cutoff is 
introduced.  In contrast, vorticity is not conserved in the quantum 
field theory because quanta can jump between modes of different 
vorticity.  The creation of vorticity is essential to explain the 
phenomenology of caustic rings and solve the galactic angular 
momentum problem \cite{Banik}.

It is pertinent, we believe, to remark that Eq. (\ref{ghp}) does 
not make sense unless a quantity with dimension of action, such 
as $\hbar$, is introduced by hand.  The classical field theory 
does not have a notion of particles nor therefore of particle 
mass, even after the wavevector cutoff $k_{\rm UV}$ has been 
introduced.  It has only modes with dispersion law
\begin{equation}
\omega(\vec{k}) = \sqrt{\omega_0^2 + c^2 \vec{k}\cdot\vec{k}}
\label{disp}
\end{equation}
where $\omega_0$ is the angular frequency of oscillation of 
the $\vec{k} = 0$ mode.  In the quantum theory, the particle 
mass is given by   
\begin{equation}
m = \omega_0 \hbar/c^2
\label{mass}
\end{equation}
but in the classical theory, with or without cutoff, there is no 
such thing as particle mass.  Likewise, Eq.~(\ref{ghp}) should 
be written 
\begin{equation}
T_{\rm crit} \sim {(\hbar k_{\rm UV})^2 \over 2m}
\label{ghp2}
\end{equation}
to be dimensionally consistent.  

\subsubsection{How long is a classical description valid?}

Granted that the outcome of thermalization in a degenerate Bosonic 
system is different from that of its classical analog, one may still 
ask how long a classical description of such a system is accurate.  This 
question was the topic of a recent paper by two of us \cite{simtherm}, 
in which it was shown by analytical arguments and numerical simulation 
of a toy model that the duration of classicality of a degenerate 
interacting Bosonic system is of order, and not longer than, its 
thermalization time $\tau$.  We summarize some results of 
ref. \cite{simtherm} here, and add toy model simulations 
that are analogous to our analytical calculations in 
Sections III and IV.

A general Bosonic system that conserves the number of quanta has a  
Hamiltonian of the form
\begin{equation}
H = \sum_j \omega_j a_j^\dagger a_j +
{1 \over 4} \sum_{jkln} \Lambda_{jk}^{ln}
a_j^\dagger a_k^\dagger a_l a_n
\label{Ham}
\end{equation}
where the $a_j$ and $a_j^\dagger$ are annihilation and creation operators 
satisfying canonical equal-time commutation relations.  One might add to the 
RHS of Eq.~(\ref{Ham}) terms of the form $a^\dagger a^\dagger a^\dagger a~a~a$, 
and so forth, but this would not alter the discussion in a significant way.
${\cal N}_j = a_j^\dagger a_j$ is the number of quanta in oscillator $j$.  
In the Heisenberg picture, the annihilation operators $a_j(t)$ satisfy 
the equations of motion
\begin{equation}
i \dot{a}_j = [a_j, H] = \omega_j a_j +       
{1 \over 2} \sum_{kln} \Lambda_{jk}^{ln} a_k^\dagger a_l a_n~~\ .
\label{Ehren}
\end{equation}             
The classical description of the system is obtained by replacing
the $a_j(t)$ with c-numbers $A_j(t)$.  They satisfy
\begin{equation}
i \dot{A}_j = \omega_j A_j +
{1 \over 2} \sum_{kln} \Lambda_{jk}^{ln} A_k^* A_l A_n~~\ .    
\label{ceom}
\end{equation}
The classical analogs of the quantum state occupation numbers 
${\cal N}_j$ are $N_j = A_j^*A_j$.  The question is: given the 
same initial values, how long do the classical analogs $N_j(t)$ 
track the expectation values $<{\cal N}_j(t)>$ of the quantum 
operators?  We call "duration of classicality" the time scale 
over which the classical description accurately describes the 
quantum system within some margin of error, say 20\%.  

To address this issue, a toy model of five quantum oscillators was 
simulated numerically \cite{simtherm}. The toy model had been previously
discussed and simulated in ref.~\cite{axtherm} to verify numerically the 
validity of formulae that estimate the rate of thermalization in the 
'condensed regime', defined by the condition $\Gamma > \delta \epsilon$
where $\Gamma$ is the thermalization rate and $\delta \epsilon$ the 
energy dispersion of the quanta in the system.   The dark matter axion 
fluid thermalizes in the condensed regime.  The Hamiltonian of the 
toy model has the form given in Eq.~(\ref{Ham}) with 
$\omega_j = j \omega_1$ ($j$ = 1, 2, 3, 4, 5) and 
$\Lambda_{jk}^{ln}  = 0$ unless $j + k = l + n$.  Non-zero values are 
given to $\Lambda_{14}^{23}$, $\Lambda_{15}^{24}$, $\Lambda_{25}^{34}$,   
$\Lambda_{22}^{13}$, $\Lambda_{33}^{24}$, $\Lambda_{33}^{15}$ and  
$\Lambda_{44}^{35}$, and their conjugates $\Lambda_{jk}^{ln} =    
\Lambda_{ln}^{jk~*}$.  The Schr\"odinger equation
\begin{equation}
i \partial_t |\Psi(t)\rangle = H |\Psi(t)\rangle        
\label{Schrod}
\end{equation}
was solved numerically for a large variety of initial conditions.  
In all cases it was found that the duration of classicality is 
less than or at most of order the relaxation time $\tau$, defined 
as the time scale over which the distribution of the quanta over 
the oscillators changes completely.  Fig. 1 shows in its top panel 
the quantum evolution of the initial state
$|{\cal N}_1, {\cal N}_2, ..., {\cal N}_5\rangle =  
|12, 25, 4, 12, 1\rangle$ as an example.  The figure shows that 
the expectation values $\langle{\cal N}_j\rangle$ move towards their 
thermal averages on the expected time scale $\tau = 1/\Gamma$, which 
is of order 0.4 given the coupling strengths $\Lambda_{jk}^{ln}$        
in the simulation \cite{axtherm}.   The thermal averages are shown 
by the dots on the right side of Fig. 1 (top panel).  The bottom 
panel of Fig. 1 shows the classical evolution of the initial state 
$(A_1, A_2, ..., A_5) = 
(\sqrt{12}, \sqrt{25}, \sqrt{4}, \sqrt{12}, \sqrt{1})$, in which   
the $N_j$ and their time derivatives $\dot{N}_j$ have the same    
initial values as their quantum analogues in the top panel.  Fig. 1 
shows that the classical evolution tracks the quantum evolution only 
for a time of order, and relatively short compared to, $\tau$.  

The toy model can be made to behave analogously to the homogeneous 
quantum field condensates discussed in Section III and IV.  The 
homogeneous condensates persist indefinitely in their classical 
description but have a finite lifetime in their quantum description.  
Initial states with the analogous property in the toy model are 
$|0,N,0,0,0\rangle$.  In their classical evolution, these states 
persist indefinitely because the RHS of Eq.~(\ref{ceom}) vanishes.  
In their quantum evolution, the quanta in the 2nd oscillator jump 
in pairs to the 1st and 3rd oscillators and thence to the 4th and 
5th oscillators.  Fig. 2 shows the $\langle{\cal N}_j(t)\rangle$ 
as a function of time for $N = 100$ in panel a), contrasted with 
the constant $N_j$ in panel b).  For the generic initial states 
simulated in ref.~\cite{simtherm}, the relaxation rate is of order 
$\Gamma \sim \Lambda \sqrt{I} {\cal N}$ for both the classical and 
quantum evolutions, where $I$ is the number of relevant interaction 
terms on the RHS of Eq.~(\ref{Ham}), and $\Lambda$ and ${\cal N}$ 
are typical values of the interaction strengths and of the quantum 
occupation numbers \cite{axtherm}.  For the special initial states 
$|0,N,0,0,0\rangle$ the relaxation rate vanishes according to the 
classical evolution but is of order 
$\Gamma \sim |\Lambda_{13}^{22}| N/\log(N)$ according to the 
quantum evolution.  The factor $\log(N)$ appears because the 
relaxation of these special states is limited by the initial 
process 2 + 2 $\rightarrow$ 1 + 3, which acts as a bottleneck. 
The 2 + 2 $\rightarrow$ 1 + 3 process causes the occupation numbers 
of the 1st and 3d oscillators to grow as $e^{|\Lambda_{13}^{22}| N t}$, 
the difference between the classical and quantum evolutions being only 
that the growth is seeded in the quantum evolution whereas it is unseeded 
in the classical evolution.  Applying the methods of Section III and IV 
to the toy model, one finds 
\begin{equation} 
\langle{\cal N}_1\rangle = \langle{\cal N}_3\rangle = \sinh^2(\gamma t) 
\label{toy} 
\end{equation} 
with $\gamma = {1 \over 2}|\Lambda_{13}^{22}| N$, and therefore 
\begin{equation} 
\langle{\cal N}_2\rangle = N - 2~\sinh^2(\gamma t) 
\label{N2} 
\end{equation} 
for $t$ sufficiently small that the condensate has not been depleted 
much yet.  The dotted lines in Fig. 2 show $\langle{\cal N}_2(t)\rangle$, 
$\langle{\cal N}_1(t)\rangle$ and  $\langle{\cal N}_3(t)\rangle$ according 
to Eqs.~(\ref{toy}) and (\ref{N2}).  The behavior of our toy model 
condensate is consistent with the discussion of a similar toy model 
condensate in ref.~\cite{Dvali13}.

In summary, the duration of classicality of the initial state 
$|0,N,0,0,0\rangle$, which persists indefinitely according to its 
classical evolution, is a factor $\log(N)$ longer than the duration 
of classicality of generic states 
$|{\cal N}_1,{\cal N}_2,{\cal N}_3,{\cal N}_4,{\cal N}_5\rangle$.  
The $|0,N,0,0,0\rangle$ state is the toy model analog of the 
homogeneous condensates discussed in Sections III and IV.  Those 
homogeneous condensates also persist forever according to their 
classical evolution, but have a finite duration of classicality 
according to their quantum evolution.  We expect the duration of 
classicality of inhomogeneous condensates to be shorter than that 
of homogeneous condensates for the same reason that the duration 
of classicality of generic toy model initial states is shorter than 
that of the $|0,N,0,0,0\rangle$ initial state, the reason being the 
absence in the case of generic states of the thermalization bottleneck 
that is present for the initial state $|0,N,0,0,0\rangle$.

\subsubsection{Homogeneity is not a necessary outcome, or criterion}

Contrary to statements appearing occasionally in the literature, 
the condensed state need not be a state of momentum $\vec{p} = 0$.  
Generally, it is not.  The state $\vec{p} = 0$ is homogeneous and 
minimizes the kinetic energy 
$\omega_{\vec{p}} = {\vec{p}\cdot\vec{p} \over 2m}$ of a particle 
in empty space.   But in general 1) space is not empty and 2) the 
particle energies $\epsilon_j$ that appear in Eq.~(\ref{Bose}) 
differ from the frequencies $\omega_j$ that appear in Eq.~(\ref{Ham}) 
because of interactions.  Even in empty space, the lowest energy 
available state need not be the zero momentum state.

Lack of homogeneity is no impediment to Bose-Einsten condensation.  
Fig. 3 shows a Bose-Einstein condensate that is highly inhomogeneous 
on some length scale $d$ but extends over a much larger length scale 
$L$.  It may be realized by placing superfluid $^4$He in a long tube 
with various obstructions on the length scale $d$ inside the tube.  
Although inhomogeneous on the length scale $d$ the condensate has 
long range correlations on the length scale $L$, as we now show 
explicitly.

For a general system undergoing Bose-Einstein condensation, let 
$u^j(\vec{x},t)$ be the wavefunction of the particle state with energy 
$\epsilon_j$.  The wavefunctions form a complete orthonormal set: 
\begin{eqnarray} 
\int_V d^3x ~ u^j(\vec{x},t)^*~u^k(\vec{x},t) &=& \delta^k_j \nonumber\\ 
\sum_j u^j(\vec{x},t)^*~u^j(\vec{y},t) &=& 
\delta(\vec{x} - \vec{y})~~\ . 
\label{ONC} 
\end{eqnarray} 
The quantum scalar field $\phi(\vec{x},t)$ describing the particles 
undergoing Bose-Einstein condensation and its canonically conjugate 
field $\pi(\vec{x},t)$ may be expanded in terms of those wavefunctions:
\begin{eqnarray} 
\phi(\vec{x},t) &=& 
\sum_j {1 \over \sqrt{2 m}}[u^j(\vec{x},t) b_j(t) + 
u^j(\vec{x},t)^* b_j(t)^\dagger]\nonumber\\ 
\pi(\vec{x},t) &=& \sum_j \sqrt{m \over 2} {1 \over i} 
[u^j(\vec{x}) b_j(t)  - u^j(\vec{x},t)^* b_j(t)^\dagger] 
\label{expand} 
\end{eqnarray} 
where the $b_j(t)$ and $b_j^\dagger(t)$ are annihilation and creation 
operators satisfying canonical equal time commutation relations.  Note 
that the $b_j(t)$ and $b_j^\dagger(t)$ in Eq.~(\ref{expand}) are different 
from the $a_j(t)$ and $a_j^\dagger(t)$ in the previous subsection since the 
latter annihilate and create particles in eigenstates of the free Hamiltonian, 
whereas the $b_j(t)$ and $b_j^\dagger(t)$ annihilate and create particles in 
eigenstates of the one-particle Hamiltonian in which the interactions of 
the one particle with all the other particles are derived from the full 
Hamiltonian using mean field theory.  

A general system state may be written 
\begin{equation} 
|\Psi\rangle = \sum_{\{{\cal N}_j\}} c(\{{\cal N}_j\}) |\{{\cal N}_j\}\rangle_t
\label{systa} 
\end{equation} 
where $\{{\cal N}_j\}$ is an arbitrary distribution of the occupation 
numbers over the particle states, 
\begin{equation} 
|\{{\cal N}_j\}\rangle_t = \prod_j {1 \over \sqrt{{\cal N}_j !}} 
(b_j(t)^\dagger)^{{\cal N}_j} |0\rangle
\label{Fockstate}
\end{equation}
and $|0\rangle$ is the empty state.   Since the total number of particles 
is conserved, we may take $|\Psi\rangle$ to be an eigenstate of the total 
number operator
\begin{equation}
\sum_j b_j(t)^\dagger b_j(t) |\Psi\rangle= N |\Psi\rangle~~\ ,
\label{eigen}
\end{equation}
in which case, $c(\{{\cal N}_j\}) = 0$ unless $\sum_j {\cal N}_j = N$.
In the state $|\Psi\rangle$, the field $\phi(\vec{x},t)$ has equal-time 
correlation function
\begin{eqnarray}
&~&\langle\Psi|\phi(\vec{x},t) \phi(\vec{y},t) |\Psi\rangle =
\sum_{\{{\cal N}_j\}} \sum_{\{{\cal N}_j^\prime\}}
c^*(\{{\cal N}_j\}) c(\{{\cal N}_j^\prime\})
\sum_{k,l} {1 \over 2m}\cdot\nonumber\\
&\cdot& 
\big[u^k(\vec{x},t)^* u^l(\vec{y},t)~ 
_t\langle\{{\cal N}_j\}| b_k^\dagger(t) b_l(t)| \{{\cal N}_j^\prime\}\rangle_t 
\nonumber\\ &+&
~u^k(\vec{x},t) u^l(\vec{y},t)^*~ 
_t\langle\{{\cal N}_j\}| b_k(t) b_l^\dagger(t)| \{{\cal N}_j^\prime\}\rangle_t\big]~\ . 
\label{corr}
\end{eqnarray}
If a Bose-Einstein condensate has formed, the lowest energy available 
state has occupation number $N_0$ of order $N$.  In that case 
$c(\{{\cal N}_j\}) = 0$ unless ${\cal N}_0 \simeq N_0$ and therefore
\begin{equation}
\langle \Psi| \phi(\vec{x},t) \phi(\vec{y},t) | \Psi \rangle
= {N_0 \over 2 m}(u^0 (\vec{x},t)^* u^0(\vec{y},t) 
+ u^0(\vec{x},t) u^0(\vec{y},t)^*) ~ +~ ... 
\label{corr2}
\end{equation}
where the dots are contributions, from particle states other than the 
condensed state, that fall off exponentially or as a power law with 
distance $|\vec{x} - \vec{y}| \equiv r$. The contribution from the 
condensed state does not fall with distance $r$.  Instead, for given 
$\vec{y}$, it has support wherever $u^0(\vec{x},t)$ has support.  Thus 
any Bose-Einstein condensate is correlated over the whole extent of 
the condensate.  

\subsubsection{What state do the particles condense into?}

Since thermalization is a condition for Bose-Einstein 
condensation, it follows that the state the particles 
condense into is the lowest energy state that is available 
to them through the thermalizing interactions.   In general 
it is not the lowest energy state in an absolute sense.   For 
example, if a beaker of superfluid $^4$He sits on a table, a 
macroscopically large number of atoms are in a condensed 
state.  This condensed state is certainly not the lowest 
energy state since its energy can be lowered by placing the 
beaker on the floor.  It is, however, the lowest energy state 
available to the $^4$He atoms through the thermalizing 
interactions.

As was already mentioned, the condensed state need not be 
stable.  Ideally, however, it ought to be stable on the 
thermalization time scale.  A complicating factor is that
thermalization is rarely complete.  Fortunately, Bose-Einstein 
condensation occurs immediately and explosively on the 
thermalization time scale.  The rate at which particles 
move to the condensed state is proportional to the number of 
particles that are already in the condensed state \cite{Semikoz}.  
The time scale over which a complete Bose-Einstein distribution 
is established is generally much longer than the time scale 
over which the Bose-Einstein condensate forms \cite{Berges}.

Nonetheless, in the case of Bose-Einstein condensation of dark 
matter axions, we have to deal with the complication that the 
axion fluid is made unstable by the very interaction that 
thermalizes it.  After Bose-Einstein condensation has occurred, 
further thermalization is required because the instability causes
the lowest energy available state to change with time.  Our paper 
is motivated by the question what is the outcome of thermalization 
while density perturbations grow and how does this outcome differ
from the predictions of cosmological perturbation theory with 
ordinary CDM.

In the remainder of this paper we construct a formalism that allows
one to discuss more clearly the thermalization of a fluid made unstable, 
and therefore inhomogeneous, by the very interactions that thermalize it.  
In Section III, we discuss the evolution of a highly degenerate Bose 
fluid with attractive $\lambda \phi^4$ interactions $(\lambda < 0$). 
In Section IV, we discuss the evolution of a highly degenerate Bose 
fluid with gravitational self-interactions.   We discuss the 
$\lambda \phi^4$ case first because it is somewhat simpler. 

\section{Attractive contact interactions}

In this section, we introduce a formalism to analyze the 
thermalization and evolution of a highly degenerate Bose
fluid with attractive contact interactions.  The interactions
play the double role of rendering the fluid unstable and of 
thermalizing it.  This section is mainly a warmup exercise 
preliminary to analyzing the thermalization and evolution 
of a highly degenerate Bose fluid with gravitational 
self-interactions in the next section.  It may also be 
useful in the analysis of some condensed matter systems.

The model we analyze is $\lambda \phi^4$ theory.  Its 
Hamiltonian is 
\begin{equation}
H = \int d^3x [{1 \over 2} (\pi)^2 + {1 \over 2} (\vec\nabla \phi)^2 
+ {1 \over 2} m^2 \phi^2 + {\lambda \over 4!} :\phi^4:] 
\label{Ham2}
\end{equation}
where $\phi(\vec{x}, t)$ and $\pi(\vec{x}, t)$ are conjugate 
Hermitian scalar fields satisfying canonical equal time commutation 
relations.  The double colon $: ... :$ symbol in the last term of 
Eq.~(\ref{Ham2}) indicates normal ordering.   That term  describes 
contact interactions.  They are repulsive if $\lambda > 0$ and 
attractive if $\lambda < 0$.  The $\phi$ and $\pi$ fields satisfy 
the equations of motion
\begin{equation}
\partial_t \phi = \pi~~~,~~~~
\partial_t \pi - \nabla^2 \phi + m^2 \phi + 
{\lambda \over 6} :\phi^3:~ = 0~~~\ .
\label{eom}
\end{equation}
We will concern ourselves only with the non-relativistic regime 
of the theory.  The non-relativistic limit is obtained by setting
\begin{eqnarray}
\phi(\vec{x}, t) &=& {1 \over \sqrt{2m}}
[\psi(\vec{x}, t) e^{- i m t} + \psi(\vec{x}, t)^\dagger e^{ i m t }]
\nonumber\\
\pi(\vec{x}, t) &=& \sqrt{m \over 2} (-i) 
[\psi(\vec{x}, t) e^{- i m t} - \psi(\vec{x}, t)^\dagger e^{ i m t }]
~~\ ,
\label{nr}
\end{eqnarray}
neglecting terms of order $\partial_t \psi$ versus terms of order
$m\psi$, and ignoring terms proportional to $e^{2 i m t}$ and 
$e^{- 2 i m t}$ that oscillate so fast in time that they effectively
average to zero.  $\psi(\vec{x}, t)$ is a non-Hermitian scalar field 
satisfying the equal time commutation relations
\begin{equation}
[\psi(\vec{x}, t), \psi(\vec{y}, t)] = 0~~~~,~~~~
[\psi(\vec{x}, t), \psi(\vec{y}, t)^\dagger] = 
\delta^3(\vec{x} - \vec{y})~~~\ ,
\label{etcm}
\end{equation}
and the equation of motion
\begin{equation}
i \partial_t \psi = - {1 \over 2m} \nabla^2 \psi 
+ {\lambda \over 8 m^2} \psi^\dagger \psi~\psi ~~\ .
\label{eom2}
\end{equation}
Note that the number of particles is conserved in the 
non-relativistic limit even though the number of particles
is not conserved in the original theory, Eq.~(\ref{Ham2}).

We expand the $\psi$ field 
\begin{equation}
\psi(\vec{x}, t) = \sum_{\vec{\alpha}} 
u^{\vec{\alpha}}(\vec{x}, t) a_{\vec{\alpha}}(t)
\label{expand2}
\end{equation}
in an orthonormal and complete set of wavefunctions 
$u^{\vec{\alpha}}(\vec{x},t)$, labeled by $\vec{\alpha}$.  Thus
\begin{eqnarray}
\int_V d^3 x ~u^{\vec{\alpha}}(\vec{x}, t)^* u^{\vec{\beta}}(\vec{x}, t) 
&=& \delta_{\vec{\alpha}}^{\vec{\beta}}~~~{\rm and}\nonumber\\ 
\sum_{\vec{\alpha}} u^{\vec{\alpha}}(\vec{x}, t)^* u^{\vec{\alpha}}(\vec{y}, t)
&=& \delta^3(\vec{x} - \vec{y})
\label{onc}
\end{eqnarray}
where $V$ is the volume of space where the theory is defined.  
The $a_{\vec{\alpha}}(t)$ and $a_{\vec{\alpha}}(t)^\dagger$ are 
annihilation and creation operators satisfying equal time 
commutation relations
\begin{equation}
[a_{\vec{\alpha}}(t), a_{\vec{\beta}}(t)] = 0~~~,~~~
[a_{\vec{\alpha}}(t), a_{\vec{\beta}}(t)^\dagger] = 
\delta_{\vec{\alpha}}^{\vec{\beta}}~~~\ ,
\label{etcm2}
\end{equation}
and the equation of motion 
\begin{equation}
i \partial_t a_{\vec{\alpha}} = 
\sum_{\vec{\beta}} M_{\vec{\alpha}}^{\vec{\beta}} a_{\vec{\beta}}
~+~ {1 \over 2} \sum_{\vec{\beta},\vec{\gamma},\vec{\delta}} 
\Lambda_{\vec{\alpha} \vec{\beta}}^{\vec{\gamma} \vec{\delta}}
~a_{\vec{\beta}}^\dagger a_{\vec{\gamma}} a_{\vec{\delta}}
\label{aeom}
\end{equation}
with 
\begin{eqnarray}
M_{\vec{\alpha}}^{\vec{\beta}}(t) &=& \int_V d^3x
~ u^{\vec{\alpha}}(\vec{x},t)^* 
(- i \partial_t - {1 \over 2m} \nabla^2) 
u^{\vec{\beta}}(\vec{x},t)
\nonumber\\
\Lambda_{\vec{\alpha} \vec{\beta}}^{\vec{\gamma} \vec{\delta}}(t) &=&
{\lambda \over 4 m^2} \int_V d^3x~u^{\vec{\alpha}}(\vec{x},t)^*
u^{\vec{\beta}}(\vec{x},t)^* u^{\vec{\gamma}}(\vec{x},t) 
u^{\vec{\delta}}(\vec{x},t)~~~\ .
\label{MLam}
\end{eqnarray}
In the non-relativistic limit, the above equations are exact for 
any orthonormal complete set of states $u^{\vec{\alpha}}$.

\subsection{Classical description}

The classical description is obtained by replacing the quantum field 
$\psi(\vec{x}, t)$ by a wavefunction $\Psi(\vec{x}, t)$.  $\Psi$ 
satisfies the c-number version of Eq.~(\ref{eom2})
\begin{equation}
i \partial_t \Psi = - {1 \over 2 m} \nabla^2 \Psi 
+ {\lambda \over 8 m^2} |\Psi|^2 \Psi 
\label{SGP}
\end{equation}
called the Schr\"odinger-Gross-Pitaevskii equation. (Although 
wavefunctions are historically associated with a quantum mechanical 
description, from the point of view of quantum field theory a 
wavefunction is merely a solution of the classical field equations
in the non-relativistic limit.)  The wavefunction may be written
\begin{equation}
\Psi(\vec{x}, t) = A(\vec{x}, t) e^{i \beta(\vec{x}, t)}~~\ ,
\label{dec}   
\end{equation}
where $A(\vec{x},t)$ and $\beta(\vec{x}, t)$ are real.  The 
wavefunction $\Psi$ describes a fluid of number density 
\begin{equation}
n(\vec{x}, t) = A(\vec{x},t)^2
\label{den}
\end{equation}
and velocity 
\begin{equation}
\vec{v}(\vec{x}, t) = {1 \over m} \vec{\nabla} \beta(\vec{x}, t)~~\ .
\label{vel}
\end{equation}
Eq.~(\ref{SGP}) implies the continuity equation
\begin{equation}
\partial_t n + \vec{\nabla}\cdot(n \vec{v}) = 0
\label{cont}
\end{equation}
and the Euler-like equation
\begin{equation}
\partial_t \vec{v} + (\vec{v} \cdot \vec{\nabla}) v = 
- {1 \over m} \vec{\nabla}V - \vec{\nabla} q
\label{Euler}
\end{equation}
where 
\begin{equation}
V(\vec{x},t) = {\lambda \over 8 m^2} n(\vec{x}, t) 
\label{pot}
\end{equation}
and  
\begin{equation}
q(\vec{x}, t) = - {1 \over 2 m^2} {\nabla^2 \sqrt{n} \over \sqrt{n}}~~\ .
\label{qp}
\end{equation}
$q(\vec{x}, t)$ is sometimes called ``quantum pressure".  Except for 
the $-\vec{\nabla} q$ term, Eq.~(\ref{Euler}) is the Euler equation 
for a fluid of classical particles moving in the potential $V$.  The 
$-\vec{\nabla} q$ term is a consequence of the underlying wave nature
of the fluid and accounts, for example, for the tendency of a wavepacket 
to spread.   

Eq.~(\ref{SGP}) admits the homogeneous solution 
\begin{equation}
\Psi_0 = \sqrt{n_0} e^{- i \delta \omega t}
\label{hom}
\end{equation}
where
\begin{equation}
\delta \omega = {\lambda n_0 \over 8 m^2}~~\ .
\label{fs}
\end{equation}
Consider small perturbations about that solution
\begin{equation}
\Psi(\vec{x}, t) = \Psi_0(t)~+~\Psi_1(\vec{x}, t)~~\ .
\label{pert}
\end{equation}
To lowest order, the perturbations satisfy
\begin{equation}
i \partial_t \Psi_1 = - {1 \over 2m} \nabla^2 \Psi_1
+ \delta \omega (2 \Psi_1 + e^{- 2 i \delta \omega t} \Psi_1^*)~~\ .
\label{peq}
\end{equation}
We decompose the perturbation in Fourier modes as follows
\begin{equation}
\Psi_1(\vec{x}, t) = e^{-i \delta \omega t}
\sum_{\vec{k}} C_{\vec{k}}(t) e^{i \vec{k} \cdot \vec{x}}~~\ .
\label{Four}
\end{equation}
The Fourier amplitudes satisfy
\begin{equation}
i \partial_t C_{\vec{k}}(t) = 
({k^2 \over 2m} + \delta \omega) C_{\vec{k}}(t)
+ \delta \omega C_{-\vec{k}}(t)^*~~\ .
\label{Aeom}
\end{equation}
The solutions may be written
\begin{equation}
C_{\vec{k}}(t) = s_{\vec{k}}(t) + r_{\vec{k}}(t)
\label{Asol}
\end{equation}
with 
\begin{equation}
s_{-\vec{k}}(t)^* = s_{\vec{k}}(t) ~~~{\rm and}~~~
r_{-\vec{k}}(t)^* = - r_{\vec{k}}(t)~~\ .
\label{parity}
\end{equation}
Eqs.~(\ref{Aeom}) imply that
\begin{equation}
r_{\vec{k}}(t) = {2mi \over k^2} \partial_t s_{\vec{k}}(t)
\label{req}
\end{equation} 
and that $s_{\vec{k}}(t)$ is a solution of 
\begin{equation}
\partial_t^2~s_{\vec{k}}(t) = 
- {k^2 \over 2m}
({k^2 \over 2m} + 2 \delta \omega) s_{\vec{k}}(t)~~\ .
\label{seq}
\end{equation}
We now discuss the repulsive ($\lambda > 0$) and attractive ($\lambda < 0$)
cases separately.

For $\lambda > 0$, the perturbations oscillate with angular frequency
\begin{equation}
\omega(k) = \sqrt{{k^2 \over 2m}
({k^2 \over 2m} + {\lambda n_0 \over 4 m^2})}~~~\ .
\label{omek}
\end{equation}
For long wavelengths, i.e. $k << k_c \equiv{\sqrt{\lambda n_0 \over 2 m}}$, 
the dispersion law is linear
\begin{equation}
\omega(k) = v_s k 
\label{lindis}
\end{equation}
with sound speed 
\begin{equation}
v_s = \sqrt{\lambda n_0 \over 8 m^3}~~\ ,
\label{sound}
\end{equation}
whereas for short wavelengths ($k >> k_c$) the dispersion law 
is quadratic.  The most general solution to Eq.~(\ref{peq}) is
Eq.~(\ref{Four}) with  
\begin{equation}
C_{\vec{k}}(t) =
c_{\vec{k}} ({k^2 \over 2m} + \omega(k)) e^{- i \omega(k) t} 
+ c_{-\vec{k}}^*({k^2 \over 2m} - \omega(k)) e^{ i \omega(k) t}
\label{pgensol}
\end{equation}
where the $c_{\vec{k}}$ are complex numbers which can be determined 
in terms of the initial perturbation $\Psi_1(\vec{x}, 0)$.

In the attractive case ($\lambda = - |\lambda|$), there is 
a critical wavelength ${2 \pi \over k_J}$, similar
to a Jeans length for gravitational interactions, with
\begin{equation}
k_J = \sqrt{|\lambda| n_0 \over 2 m}~~\ .
\label{Jlength}
\end{equation}
For $k > k_J$, the perturbations oscillate with angular 
frequency 
\begin{equation}
\omega(k) = \sqrt{{k^2 \over 2m}
({k^2 \over 2m} - {|\lambda| n_0 \over 4 m^2})}~~\ .
\label{freq}
\end{equation}
They are described by the same equations as in the previous 
paragraph.

For $k < k_J$, the perturbations are unstable.   They grow and 
decay at the rate
\begin{equation}
\gamma(k) = \sqrt{{k^2 \over 2m}
({|\lambda| n_0 \over 4 m^2} - {k^2 \over 2m})}
= {k \over 2m}\sqrt{k_J^2 - k^2}~~\ .
\label{rate}
\end{equation}
The solutions to Eq.~(\ref{Aeom}) are 
\begin{equation}
C_{\vec{k}}(t) =
({k^2 \over 2m} + i \gamma(k)) c_{\vec{k},+} e^{\gamma(k) t}
+ ({k^2 \over 2m} - i \gamma(k)) c_{\vec{k},-} e^{- \gamma(k) t}
\label{insA}
\end{equation}
where $c_{\vec{k},\pm}$ are complex numbers subject to the constraint 
$c_{\vec{k},\pm} = c_{-\vec{k},\pm}^*$.  The instability occurs because 
the attractive contact forces produce a tendency of quanta to move towards 
regions of high density, and hence crowd places that are already crowded.  
This tendency overcomes the effect of quantum pressure on length scales 
larger than $k_J^{-1}$.  

\subsection{Quantum evolution}

Consider a particular solution of the classical equations of motion.
We ask:  for how long does it provide an accurate description of the 
quantum system?  A solution $\Psi(\vec{x}, t)$ of the Schr\"odinger-
Gross-Pitaevskii equation (\ref{SGP}) is a particular mode of the 
quantum field.  When this and only this mode is highly occupied, the 
classical description is very accurate, corrections being of order 
$1/N$ where $N$ is the occupation number of the mode.  The question 
is whether the quanta stay in the mode $\Psi(\vec{x}, t)$.  And, if 
they do not stay in that mode, what is the rate at which they leave it.  

To address these issues we introduce a set of modes that are 
similar to $\Psi(\vec{x}, t)$ but differ from it by long wavelength 
modulations:
\begin{equation}
u^{\vec{k}}(\vec{x}, t) = {1 \over \sqrt{N}} 
\Psi(\vec{x}, t) e^{i \vec{k}\cdot\vec{\chi}(\vec{x}, t)}~~\ .
\label{Pmodes}
\end{equation}
The $\vec{\chi}(\vec{x},t)$ are co-moving coordinates chosen 
so that the density in $\vec{\chi}$-space is constant in 
both space and time.  Thus
\begin{equation}
{d^3 N \over d \chi^3} = {n(\vec{x}, t) \over J(\vec{x}, t)} = n_0 
\label{def}
\end{equation}
where $n_0$ is a constant, $n(\vec{x}, t)$ is the physical space
density implied by $\Psi(\vec{x}, t)$ [Eq.~(\ref{den})] and 
\begin{equation}
J(\vec{x}, t) = 
| \det \left({\partial \vec{\chi} \over \partial \vec{x}}\right)|
\label{Jac}
\end{equation}
is the Jacobian of the map.  The $\vec{\chi}(\vec{x},t)$ can be 
constructed as follows.  The wavefunction $\Psi(\vec{x}, t)$ 
implies a velocity field $\vec{v}(\vec{x}, t)$, given by 
Eq.~(\ref{vel}), and hence a map $\vec{x} (\vec{\chi}, t)$:
\begin{equation}
{\partial \vec{x} \over \partial t}\Big|_{\vec{\chi}} = 
\vec{v}(\vec{x}(\vec{\chi}, t), t)~~\ . 
\label{flow}
\end{equation}
If $\vec{v}(\vec{x}, t)$ were the velocity field of a flow
of particles, $\vec{\chi}$ would label individual particles
in the flow.  For example, $\vec{\chi}$ may be the position 
of the particle at some initial time $t_*$. The map 
$\vec{\chi}(\vec{x},t)$ is the inverse of $\vec{x} (\vec{\chi}, t)$.  
This construction ensures that the density in $\vec{\chi}$-space 
is time-independent.  Furthermore it is always possible to change 
variables $\vec{\chi} \rightarrow \vec{\chi}^{\,\prime}$ such 
that the density in $\vec{\chi}^{\,\prime}$-space is 
$\vec{\chi}^{\,\prime}$-independent as well.

We choose the region in $\vec{\chi}$-space where the theory is 
defined to be a cube of volume $V_0 = L_0^3$ with periodic 
boundary conditions at its surface.   Thus the wavevectors
appearing in Eq.~(\ref{Pmodes}) are 
$\vec{k} = {2 \pi \over L_0}(n_1, n_2, n_3)$ where the 
$n_j = 0, \pm 1, \pm 2, ...$ ($j = 1,2, 3$).  We have then
\begin{equation}
\int_V d^3x~u^{\vec{k}}(\vec{x}, t)^* 
u^{\vec{k}^\prime} (\vec{x}, t) = 
{1 \over V_0} \int_{V_0} d^3\chi~ 
e^{i (\vec{k}^\prime - \vec{k})\cdot\vec{\chi}} 
= \delta_{\vec{k}}^{\vec{k}^\prime}~~\ , 
\label{ONC2}
\end{equation}
i.e. the $u^{\vec{k}}(\vec{x}, t)$ form a complete 
orthonormal set. We expand the quantum field
\begin{equation}
\psi(\vec{x}, t) = \sum_{\vec{k}} 
u^{\vec{k}}(\vec{x}, t) a_{\vec{k}}(t)~~\ .
\label{exp}
\end{equation}
Eqs.~(\ref{etcm2}) - (\ref{MLam}) apply with the indices
$\vec{\alpha}$ replaced by the $\vec{\chi}$-space wavevectors 
$\vec{k}$.  Substituting Eq.~(\ref{Pmodes}), we have
\begin{eqnarray}
\Lambda_{\vec{k}_1 \vec{k}_2}^{\vec{k}_3 \vec{k}_4}(t) &=&
{\lambda \over 4 m^2 N^2} \int_V d^3x~(n(\vec{x}, t))^2 
e^{i(\vec{k}_3 + \vec{k}_4 - \vec{k}_1 - \vec{k}_2)
\cdot\vec{\chi}(\vec{x}, t)} \nonumber\\
&=& {\lambda \over 4 m^2 N}
\tilde{n}(\vec{k}_1 + \vec{k}_2 - \vec{k}_3 - \vec{k}_4, t)
\label{Lam} 
\end{eqnarray}
where
\begin{equation}
\tilde{n}(\vec{q}, t) = {1 \over V_0} \int_{V_0}
d^3\chi~n(\vec{x}(\vec{\chi}, t), t) e^{- i \vec{q}\cdot\vec{\chi}}~~\ , 
\label{ntilde}
\end{equation}
and 
\begin{eqnarray}
M_{\vec{k}}^{\vec{k}^\prime}(t) &=& 
- {\lambda \over 8 m^2} \tilde{n}(\vec{k} - \vec{k}^\prime,t)
\nonumber\\
&+&{1 \over 2mN} \int_V d^3x~n(\vec{x}, t) 
\vec{\nabla}(\vec{k}\cdot\vec{\chi}(\vec{x}, t)) \cdot
\vec{\nabla}(\vec{k}^\prime\cdot\vec{\chi}(\vec{x}, t))
e^{i(\vec{k}^\prime - \vec{k})\cdot\vec{\chi}(\vec{x}, t)}~~\ .
\label{M2}
\end{eqnarray}
The somewhat lengthy derivation of Eq.~(\ref{M2}) is given 
in the Appendix.  

The annihilation operators satisfy the equation of motion
\begin{equation}
i \partial_t a_{\vec{k}} = \sum_{\vec{k}^\prime} 
M_{\vec{k}}^{\vec{k}^\prime} a_{\vec{k}^\prime}
+ {1 \over 2} \sum_{\vec{k}_2,\vec{k}_3,\vec{k}_4}
\Lambda_{\vec{k} \vec{k}_2}^{\vec{k}_3 \vec{k}_4}
a_{\vec{k}_2}^\dagger a_{\vec{k}_3} a_{\vec{k}_4}~~\ .
\label{eom3}
\end{equation}
The corresponding classical equations of motion are
\begin{equation}
i \partial_t A_{\vec{k}} = \sum_{\vec{k}^\prime}
M_{\vec{k}}^{\vec{k}^\prime} A_{\vec{k}^\prime}
+ {1 \over 2} \sum_{\vec{k}_2,\vec{k}_3,\vec{k}_4}
\Lambda_{\vec{k} \vec{k}_2}^{\vec{k}_3 \vec{k}_4}
A_{\vec{k}_2}^* A_{\vec{k}_3} A_{\vec{k}_4}~~\ .
\label{clm3}
\end{equation}
The classical solution with which we started is 
$\Psi(\vec{x}, t) = \sqrt{N} u^{\vec{0}}(\vec{x}, t)$. 
Therefore 
\begin{equation}
A_{\vec{k}}(t) = \sqrt{N} \delta_{\vec{k}}^{\vec{0}}
\label{clsol}
\end{equation}
must solve Eq.~(\ref{clm3}).  Since
\begin{equation}
M_{\vec{k}}^{\vec{0}}(t) = 
- {N \over 2} \Lambda_{\vec{k} \vec{0}}^{\vec{0} \vec{0}}(t)
\label{const}
\end{equation}
one can verify that this is indeed the case.  Eq.~(\ref{const})
provides a consistency check on our formalism.

\subsection{Bogoliubov's quasi-particles}

Let us apply our formalism to the homogeneous state in the repulsive 
case.   We will be following in the footsteps of N.N. Bogoliubov's 
famous 1947 paper \cite{Bog}.  The homogeneous state is described by 
Eqs.~(\ref{hom}) and (\ref{fs}). Since $\vec{v} = 0$ in this state, we 
choose $\vec{\chi} = \vec{x}$ and hence
\begin{eqnarray}
u^{\vec{k}}(\vec{x}, t) &=& {1 \over \sqrt{V}} e^{- i \delta \omega t
+ i \vec{k}\cdot\vec{x}}\nonumber\\
M_{\vec{k}}^{\vec{k}^\prime} &=& ({k^2 \over 2 m} - \delta \omega)
\delta_{\vec{k}}^{\vec{k}^\prime}\nonumber\\
\Lambda_{\vec{k}_1 \vec{k}_2}^{\vec{k}_3 \vec{k}_4} &=&
{2 \delta \omega \over N} 
\delta_{\vec{k}_1 + \vec{k}_2}^{\vec{k}_3 + \vec{k}_4}~~\ .
\label{homc}
\end{eqnarray}
The equations of motion for the annihilation operators are 
therefore
\begin{equation}
i \partial_t a_{\vec{k}} = ({k^2 \over 2 m} - \delta \omega)a_{\vec{k}}
+ {\delta \omega \over N} \sum_{\vec{k}_1 \vec{k}_2}
a^\dagger_{\vec{k}_1 + \vec{k}_2 - \vec{k}} a_{\vec{k}_1} a_{\vec{k}_2}~~\ .
\label{homeom}
\end{equation}
To analyze the behavior of the system when the homogeneous 
particle state is occupied by a huge number $N$ of quanta, we 
substitute
\begin{equation}
a_{\vec{k}}(t) = \sqrt{N} \delta_{\vec{k}}^{\vec{0}}
+ b_{\vec{k}}(t)~~\ . 
\label{sub}
\end{equation}
The $b_{\vec{k}}(t)$ operators satisfy canonical commutation
relations, and the equations of motion for $\vec{k} \neq 0$ 
\begin{eqnarray}
i \partial_t b_{\vec{k}} &=& 
({k^2 \over 2m} + \delta \omega) b_{\vec{k}}
+ \delta \omega b_{-\vec{k}}^\dagger
\nonumber\\
&+& {\delta \omega \over \sqrt{N}} \sum_{\vec{k}^\prime} 
(b_{\vec{k}-\vec{k}^\prime} b_{\vec{k}^\prime}
+ 2 b_{\vec{k}^\prime-\vec{k}}^\dagger b_{\vec{k}^\prime})
+ {\delta\omega \over N} \sum_{\vec{k}_1 \vec{k}_2}
b_{\vec{k}_1 + \vec{k}_2 - \vec{k}}^\dagger b_{\vec{k}_1} b_{\vec{k}_2}~~\ .
\label{beom}
\end{eqnarray}
The last two terms describe interactions since they are respectively 
quadratic and cubic in the $b_{\vec{k}}$'s.  The interactions are 
suppressed relative to the linear terms by one or two factors of 
$1/\sqrt{N}$. 

Ignoring interactions for the time being we have 
\begin{equation}
i \partial_t \left( \begin{array} {c} b_{\vec{k}} 
\\ b_{-\vec{k}}^\dagger \end{array}\right) = 
\left( \begin{array} {cc} {k^2 \over 2m} + \delta\omega~~ & ~~\delta\omega \\
- \delta\omega ~~ &~~ - {k^2 \over 2m} - \delta\omega \end{array} \right)
\left( \begin{array} {c} b_{\vec{k}} 
\\ b_{-\vec{k}}^\dagger \end{array}\right) ~~\ .
\label{ignint}
\end{equation}
We diagonalize the matrix appearing in Eq.~({\ref{ignint}) by a 
Bogoliubov transformation:
\begin{equation}
\left( \begin{array} {c} b_{\vec{k}} 
\\ b_{-\vec{k}}^\dagger \end{array}\right) =
\left( \begin{array} {cc} u ~~ & ~~v\\
v ~~ &~~ u \end{array} \right)
\left( \begin{array} {c} \beta_{\vec{k}}
\\ \beta_{-\vec{k}}^\dagger \end{array}\right)
\label{bov}
\end{equation}
where $u$ and $v$ are real and $u^2 - v^2 = 1$.  The transformation 
from the $b_{\vec{k}}$ to the $\beta_{\vec{k}}$ is canonical.  We may 
write $u = \cosh(\eta)$ and $v = \sinh(\eta)$.  The new operators satisfy 
the equations of motion
\begin{eqnarray}
i \partial_t \left( \begin{array} {c} \beta_{\vec{k}}
\\ \beta_{-\vec{k}}^\dagger \end{array}\right)~~\ &=& 
\left( \begin{array} {cc} u ~~ & ~~ - v\\
- v ~~ &~~ u \end{array} \right) 
\left( \begin{array} {cc} a~~ & ~~b \\
- b ~~ &~~ - a \end{array} \right)
\left( \begin{array} {cc} u ~~ & ~~ v\\
v ~~ &~~ u \end{array} \right)
\left( \begin{array} {c} \beta_{\vec{k}}
\\ \beta_{-\vec{k}}^\dagger \end{array}\right) 
\nonumber\\
&=& \left( \begin{array} {cc} 
(u^2 + v^2) a + 2 u v b ~~ & ~~ 2 u v a + (u^2 + v^2) b\\
- 2 u v a - (u^2 + v^2) b ~~ &~~ - (u^2 + v^2) a - 2 u v b 
\end{array} \right)
\left( \begin{array} {c} \beta_{\vec{k}}
\\ \beta_{-\vec{k}}^\dagger \end{array}\right)
\label{liu}
\end{eqnarray}
where
\begin{equation}
a = {k^2 \over 2m} + \delta \omega ~~~~{\rm and}~~~~~
b = \delta \omega~~\ .
\label{def2}
\end{equation}
The matrix that appears on the RHS of Eq.~(\ref{liu}) is 
diagonal when
\begin{equation}
{2 u v \over u^2 + v^2} = \tanh(2 \eta) = - {b \over a}~~\ ,
\label{set}
\end{equation}
with the magnitude of the diagonal elements equal to
\begin{equation}
(u^2 + v^2) a + 2 u v b = \sqrt{a^2 - b^2} = 
\sqrt{{k^2 \over 2m}({k^2 \over 2m} + {\lambda n_0 \over 4 m^2})}
= \omega(k)
\label{omag}
\end{equation}
where $\omega(k)$ is the angular oscillation frequency that 
appears in the classical description. With this choice of $\eta$ 
\begin{equation}
i \partial_t \beta_{\vec{k}} = \omega(k) \beta_{\vec{k}} + ...
\label{go}
\end{equation}
where the dots represent interaction terms.  The Hamiltonian 
for the $\beta_{\vec{k}}$ is thus
\begin{equation}
H = \sum_{\vec{k}} \omega(k) \beta_{\vec{k}}^\dagger \beta_{\vec{k}}~~
+ H_{\rm int}
\label{Ham3}
\end{equation}
The $\beta_{\vec{k}}^\dagger$ and $\beta_{\vec{k}}$ create and annihilate 
``quasi-particles".  The quasi-particles are the quanta of excitation of 
the system when the homogeneous particle state ($\vec{k} = 0$) is hugely 
occupied.

\subsection{Instability by parametric resonance}

We now turn to the attractive case, which is of greater interest to us 
because of its analogy with gravity.  In the classical description of 
small perturbations to the homogeneous condensate with $\lambda \phi^4$
interactions, one goes from the stable repulsive case to the unstable 
attractive case by merely changing the sign of $\lambda$.  For $\lambda < 0$
and $k < k_J = \sqrt{|\lambda| n_0 \over 2m}$, $\omega(k)$ is imaginary 
and the instability is that of inverted harmonic oscillators.  Let 
us see how the instability manifests itself in the quantum description.

Eqs.~(\ref{homc}) - (\ref{def2}) are still valid when $\lambda < 0$.  For 
$k > k_J$, we perform the same steps as in Eqs.~(\ref{set}) - (\ref{go}). 
Thus for $k > k_J$, there is a set of quasi-particles as before.  For 
$k < k_J$, it is not possible to satisfy Eq.~(\ref{set}) because $|b| > |a|$. 
For $k < k_J$, we set the {\it diagonal} elements in the matrix on the RHS of 
Eq.~(\ref{liu}) equal to zero by choosing $\eta$: 
\begin{equation}
{2 u v \over u^2 + v^2} = \tanh (2 \eta) = - {a \over b}~~\ .
\label{set2}
\end{equation}
The magnitude of the off-diagonal elements is then
\begin{equation}
2 v u a + (u^2 + v^2) b = - \sqrt{b^2 - a^2} = 
- {k \over 2m} \sqrt{k_J^2 - k^2} = - \gamma(k)
\label{gk}
\end{equation}
where $\gamma(k)$ is the rate of instability that appears in the classical 
description.  With this choice of $\eta$
\begin{equation}
i \partial_t \beta_{\vec{k}} = - \gamma(k) \beta_{-\vec{k}}^\dagger
+ ...
\label{goi}
\end{equation}
where the dots represent interaction terms.  The Hamiltonian for 
the attractive case is thus
\begin{equation}
H = \sum_{\vec{k} \atop k > k_J} \omega(k) \beta_{\vec{k}}^\dagger \beta_{\vec{k}}
+ \sum_{\vec{k} \atop k < k_J , k_z > 0} 
(- \gamma(k)) (\beta_{\vec{k}} \beta_{-\vec{k}} +
\beta_{\vec{k}}^\dagger \beta_{-\vec{k}}^\dagger) + H_{\rm int}~~\ .
\label{Ham4}
\end{equation}
We may rewrite the kinetic terms for the $k < k_J$ modes  
\begin{equation}
- \gamma(k)(\beta_{\vec{k}} \beta_{-\vec{k}}
+ \beta_{\vec{k}}^\dagger \beta_{-\vec{k}}^\dagger) = 
 - {1 \over 2} \gamma(k) (\alpha_{\vec{k}}\alpha_{\vec{k}} +
\alpha_{\vec{k}}^\dagger \alpha_{\vec{k}}^\dagger)
+ {1 \over 2} \gamma(k) (\alpha_{\vec{k}}^\prime\alpha_{\vec{k}}^\prime +
\alpha_{\vec{k}}^{\prime~\dagger} \alpha_{\vec{k}}^{\prime ~\dagger})
\label{Ham5}
\end{equation}
where 
\begin{equation}
\alpha_{\vec{k}} = {1 \over \sqrt{2}}(\beta_{\vec{k}} + \beta_{-\vec{k}})
~~~{\rm and}~~~\alpha_{\vec{k}}^\prime = 
{1 \over \sqrt{2}}(\beta_{\vec{k}} - \beta_{-\vec{k}})~~\ .
\label{finop}
\end{equation}
We now show that this system exhibits parametric resonance. 

Consider the Hamiltonian for any of the $k < k_J$ modes
\begin{equation}
H = {1 \over 2}\gamma(\alpha\alpha + \alpha^\dagger \alpha^\dagger)
\label{Ham6}
\end{equation}
where $\gamma$ is real and $[\alpha,\alpha^\dagger] = 1$.  
Eq.~(\ref{Ham6}) implies the equation of motion 
\begin{equation}
i \partial_t \alpha = \gamma \alpha^\dagger~~\ .
\label{eom6}
\end{equation}
Its most general solution is
\begin{equation}
\alpha(t) = e^{iHt} \alpha(0) e^{-iHt} =
\cosh(\gamma t) \alpha(0) 
- i \sinh(\gamma t) \alpha(0)^\dagger~~~\ .
\label{sol6}
\end{equation}
The exponential growth of $\alpha(t)$ implies an instability.  
To see its implications, we go to the Schr\"odinger picture and 
obtain the evolution $|\psi(t)\rangle = e^{-iHt}|\psi(0)\rangle$ of 
the initial state $|\psi(0)\rangle = |0\rangle$, defined by 
\begin{equation}
\alpha(0)|0\rangle = 0~~\ .
\label{vac}
\end{equation}
We expand
\begin{equation}
|\psi(t)> = \sum_{n=0}^\infty c_n(t) |n\rangle
\label{exp2}
\end{equation}
where
\begin{equation}
|n\rangle = {1 \over \sqrt{n!}} (\alpha(0)^\dagger)^n |0\rangle~~\ .
\label{nstate}
\end{equation}
Eq.~(\ref{vac}) implies 
\begin{equation}
\alpha(-t) |\psi(t)\rangle = 0~~\ .
\label{revac}
\end{equation}
Combining Eqs.~(\ref{exp2}) and (\ref{revac}) one finds the
recursion relation
\begin{equation}
c_{n+1}(t) = - i \tanh(\gamma t) \sqrt{n \over n+1} c_{n-1}(t)~~\ .
\label{recur}
\end{equation}
Since $H$ changes the occupation number $n$ by $\pm 2$ only, 
$c_n(t) = 0$ for all odd $n$.  For $n = 2p$, Eq.~(\ref{recur}) 
implies
\begin{equation}
c_{2p}(t) = (- i \tanh(\gamma t))^p 
\sqrt{(2p-1)!! \over 2^p~p!} c_0(t)~~\ .
\label{2psol}
\end{equation}
The normalization condition
\begin{equation}
\sum_{p=0}^\infty~|c_{2p}(t)|^2 = 1
\label{norm}
\end{equation}
yields 
\begin{equation}
c_0(t) = {1 \over \sqrt{\cosh(\gamma t)}}~~\ .
\label{gramp}
\end{equation}
The probabilty that the system is found in the $(2p)^{\rm th}$ excited
state is thus 
\begin{equation}
|c_{2p}(t)|^2 = {(2p-1)!! \over 2^p~p!} 
{(\tanh(\gamma t))^{2p} \over \cosh(\gamma t)}~~\ .
\label{2pprop}
\end{equation}
The average occupation number can be obtained directly from
Eq.~(\ref{sol6}): 
\begin{equation}
\langle N(t) \rangle = 
\langle 0| \alpha^\dagger(t) \alpha(t) |0\rangle
= \sinh^2(\gamma t)~~\ .
\label{avN}
\end{equation}
Likewise, the average occupation number squared
\begin{equation}
\langle N(t)^2 \rangle = 
\langle 0| \alpha^\dagger(t) \alpha(t) 
\alpha^\dagger(t) \alpha(t) |0 \rangle = 
\sinh^4(\gamma t) + 
2 \cosh^2 (\gamma t) \sinh^2(\gamma t)~~\ .
\label{avN2}
\end{equation}
The root mean average deviation from the average occupation 
number is therefore
\begin{equation}
\delta N(t) = {1 \over \sqrt{2}} |\sinh(2 \gamma t)|~~\ .
\label{rmsd}
\end{equation}
Both the average occupation number and its root mean 
square deviation grow as $e^{2 \gamma t}$.  

\subsection{Duration of classicality}

We found that the occupation numbers of all $\alpha_{\vec{k}}$ 
and $\alpha^\prime_{\vec{k}}$ modes in the wavevector range 
$0 < k < k_J$ grow exponentially with rate 
$2 \gamma(k) = {k \over m} \sqrt{k_J^2 - k^2}$ at the expense 
of the $\vec{k} = 0$ condensate.  The interactions cause quanta 
to jump, two at the time, out of the condensate into the $k < k_J$ 
modes.  In contrast, the classical solution 
$A_{\vec{k}}(t) = \sqrt{N} \delta_{\vec{k}}^{\vec{0}}$ is valid 
for all time.  The $b_{\vec{k}}(t)$ annihilation operators (for 
$\vec{k} \neq 0$) are given in terms of the $\alpha_{\vec{k}}(0)$ 
and the $\alpha^\prime_{\vec{k}}(0)$ by 
\begin{eqnarray}
b_{\vec{k}}(t) &=& 
{u \over \sqrt{2}}(\alpha_{\vec{k}}(t) + \alpha^\prime_{\vec{k}}(t))
+ {v \over \sqrt{2}}(\alpha_{\vec{k}}^\dagger(t) 
- \alpha_{\vec{k}}^{\prime~\dagger}(t))
\nonumber\\
&=&{1 \over \sqrt{2}} (\alpha_{\vec{k}}(0) + \alpha^\prime_{\vec{k}}(0))
\left[u \cosh(\gamma(k)t) - i v \sinh(\gamma(k)t)\right] 
\nonumber\\
&+& {1 \over \sqrt{2}} (\alpha_{\vec{k}}^\dagger(0) 
- \alpha_{\vec{k}}^{\prime~\dagger}(0)) 
\left[v \cosh(\gamma(k)t) + i u \sinh(\gamma(k)t)\right]
\label{orig}
\end{eqnarray}
where we used Eqs.~(\ref{finop}), and Eq.~(\ref{sol6}) with 
$\gamma = \pm \gamma(k)$.  The discussion in the previous subsection 
suggests that the quantum state with the longest duration of classicality 
for describing the homogeneous condensate is $|\Psi \rangle$:
\begin{equation}
\alpha_{\vec{k}}(0) |\Psi \rangle = 
\alpha^\prime_{\vec{k}}(0) |\Psi \rangle = 0
\label{state}
\end{equation}
for all $\vec{k} \neq 0$.  Using Eq.~(\ref{orig}) we find that the average
number of quanta that have jumped from the condensate into mode $b_{\vec{k}}$, 
with $0 < k < k_J$, is 
\begin{equation}
\langle \Psi| b_{\vec{k}}^\dagger(t)  b_{\vec{k}}(t) | \Psi \rangle =
\sinh^2(\eta) + 
\cosh(2 \eta) \sinh^2(\gamma(k) t)
\label{Nkt}
\end{equation}
in state $|\Psi \rangle$ at time $t$.  As an alternative to $|\Psi\rangle$, 
we also considered the quantum state $|\Psi^\prime \rangle$ defined by:
\begin{equation}
b_{\vec{k}}(0) |\Psi^\prime \rangle = 0
\label{alstate}
\end{equation}
for all $\vec{k} \neq 0$.  Using Eq.~(\ref{orig}) we find 
\begin{equation}
\langle \Psi^\prime| b_{\vec{k}}^\dagger(t)  b_{\vec{k}}(t) | \Psi^\prime \rangle 
= \cosh^2(2 \eta) \sinh^2(\gamma(k) t)~~\ .
\label{alNkt}
\end{equation}
Although in state $|\Psi^\prime \rangle$ all $\vec{k} \neq 0$ oscillators 
are empty initially, they end up more highly occupied at later times than 
in state $|\Psi \rangle$.   The latter state has the longer duration of 
classicality, and is the state that we consider henceforth.

The average number of quanta that have evaporated from the $\vec{k} = 0$ 
condensate is 
\begin{eqnarray}
N_{\rm ev}(t) &=& \sum_{\vec{k}} 
\langle \Psi | b_{\vec{k}}^\dagger (t) b_{\vec{k}}(t) | \Psi \rangle
\nonumber\\ 
&=&  V \int_{k < k_J} {d^3 k \over (2 \pi)^3} 
{k_J^2  \over 8 k \sqrt{k_J^2 - k^2}} e^{2 \gamma(k) t}
= {V k_J^3 \over 16 \pi^2} \int_0^1 du~{u \over \sqrt{1 - u^2}}
~e^{{k_J^2 t \over m} u \sqrt{1 - u^2}}
\nonumber\\&\simeq& 
{V k_J^3 \over 16 \pi^2} \sqrt{\pi m \over 2 k_J^2 t} e^{k_J^2 t \over 2 m}
= {N |\lambda| \over 32 \pi \sqrt{2 \pi m t}} e^{k_J^2 t \over 2 m}
\label{Nprime}
\end{eqnarray}
in state $|\Psi \rangle$ for $t >> m/k_J^2$.   We used the saddle point 
approximation in the 
penultimate step.  We thus find that after a time of order 
\begin{equation}
t_c \sim {2 m \over k_J^2} 
\ln\left({32 \pi^{3 \over 2} n_0 \over k_J^3}\right)
\label{cdur}
\end{equation}
the $\vec{k} = 0$ condensate is almost entirely depleted. Up to 
a numerical constant, the argument of the logarithm in Eq.~(\ref{cdur})
is the number of quanta in a sphere of radius $1/k_J$.

\section{Gravitational self-interactions}

We now come to our main topic: the dynamics of a degenerate quantum 
scalar field interacting with itself through Newtonian gravity.  We 
must be in the non-relativistic regime of this system for Newtonian 
gravity to be valid, i.e. only slow moving ($v << c$) quanta are 
excited.  The dynamics is in terms of the non-Hermitian scalar field 
$\psi(\vec{r}, t)$ introduced in Eqs.~(\ref{nr}).  $\psi(\vec{r}, t)$ 
satisfies the equal time commutation relations (\ref{etcm}).  The 
Hamiltonian is 
\begin{eqnarray}
H &=& \int_V d^3r~{1 \over 2m} \vec{\nabla} \psi(\vec{r}, t)^\dagger
\cdot \vec{\nabla} \psi(\vec{r}, t)\nonumber\\
&-& {G m^2 \over 2} \int_V d^3r \int_V d^3r^{\,\prime}
{1 \over |\vec{r} - \vec{r}^{\,\prime}|}
\psi(\vec{r}, t)^\dagger \psi(\vec{r}, t) 
\psi(\vec{r}^{\,\prime}, t)^\dagger \psi(\vec{r}^{\,\prime}, t)~~\ .
\label{Ham10}
\end{eqnarray}
The equation of motion is
\begin{equation}
i \partial_t \psi(\vec{r}, t) = 
[\psi(\vec{r}, t), H] = - {1 \over 2m} \nabla^2 \psi(\vec{r}, t)
+ m \varphi(\vec{r}, t) \psi(\vec{r}, t) 
\label{eom10}
\end{equation}
where $\varphi(\vec{r}, t)$ is the operator
\begin{equation}
\varphi(\vec{r}, t) = - G m \int_V d^3r^{\,\prime}~
{\psi(\vec{r}^{\,\prime}, t)^\dagger \psi(\vec{r}^{\,\prime}, t) 
\over |\vec{r} - \vec{r}^{\,\prime}|}
\label{opN}
\end{equation}
whose classical analog is the gravitational potential.  
We may expand $\psi(\vec{r}, t)$ in any orthonormal and 
complete set of wavefunctions $u^{\vec{\alpha}}(\vec{r}, t)$, 
as we did in Section III.  Eqs.~(\ref{expand2} - \ref{MLam})
remain unchanged except that
\begin{eqnarray}
\Lambda_{\vec{\alpha} \vec{\beta}}^{\vec{\gamma} \vec{\delta}}
&=& - G m^2 \int_V d^3r \int_V d^3r^{\,\prime} 
{1 \over |\vec{r} - \vec{r}^{\,\prime}|}\cdot\nonumber\\
&\cdot&u^{\vec{\alpha}}(\vec{r}, t)^* u^{\vec{\beta}}(\vec{r}^{\,\prime}, t)^*
(u^{\vec{\gamma}}(\vec{r}, t) u^{\vec{\delta}}(\vec{r}^{\,\prime}, t) +
u^{\vec{\gamma}}(\vec{r}^{\,\prime}, t) u^{\vec{\delta}}(\vec{r}, t))
\label{gLam}
\end{eqnarray}
is substituted for the second equation (\ref{MLam}).

\subsection{Classical description}

In the classical description, the operator $\psi(\vec{r}, t)$ is replaced
by a c-number wavefunction $\Psi(\vec{r}, t)$.   The wavefunction satisfies
the classical analog of Eq.~(\ref{eom10}):
\begin{equation}
i \partial_t \Psi(\vec{r}, t) = - {1 \over 2m} \nabla^2 \Psi(\vec{r}, t) 
+ m \Phi(\vec{r}, t) \Psi(\vec{r}, t)
\label{eom11}
\end{equation}
with 
\begin{equation}
\Phi(\vec{r}, t) = - G m \int_V d^3r^{\,\prime}~
{|\Psi(\vec{r}^{\,\prime}, t)|^2 \over |\vec{r} - \vec{r}^{\,\prime}|}~~\ .
\label{Newpot}
\end{equation}
The gravitational potential $\Phi(\vec{r}, t)$ satisfies the Poisson 
equation
\begin{equation}
\nabla^2 \Phi(\vec{r}, t) = 4 \pi G m |\Psi(\vec{r}, t)|^2~~\ .
\label{Poiss}
\end{equation}
Let us remark however that Eq.~(\ref{Poiss}) implies Eq.~(\ref{Newpot}) 
only up to a solution of the Laplace equation.  The Schr\"odinger-Poisson 
equations, Eqs.~(\ref{eom11}) and (\ref{Poiss}), are commonly used to 
describe self-gravitating degenerate axions or axion-like particles 
\cite{ULALP}.  They were used in ref.~\cite{Christ} to describe the 
homogeneous expanding universe and the evolution of density perturbations 
therein.  We summarize the results of ref. \cite{Christ} as they are the 
starting point for our analysis of the system's quantum evolution in the 
next subsection.

The wavefunction that describes the homogeneous expanding universe is 
\begin{equation} 
\Psi_0(\vec{r}, t) = \sqrt{n_0(t)} e^{i {1 \over 2} m H(t) r^2} 
\label{homun} 
\end{equation} 
where $H(t)$ is the Lema\^itre-Hubble expansion rate.  Indeed the 
velocity field implied by Eq.~(\ref{homun}) is 
\begin{equation} 
\vec{v}(\vec{r}, t) = H(t) \vec{r}~~\ . 
\label{LH} 
\end{equation} 
Furthermore, Eqs.~(\ref{eom11}) and (\ref{Poiss}) imply the 
continuity equation
\begin{equation}
\partial_t n_0 + 3 H n_0 = 0
\label{coneq}
\end{equation}
and the Friedmann equation
\begin{equation}
H(t)^2 + {K \over a(t)^2} = {8 \pi G \over 3} m n_0(t)~~\ .
\label{fried}
\end{equation}
$K = +1, 0, -1$ depending on whether the universe is closed, 
critical or open and $a(t)$ is the scale factor defined by 
$H(t) = {\dot{a} \over a}$.  Eqs.~(\ref{LH}), (\ref{coneq}) 
and (\ref{fried}) are the standard equations that describe 
the homogeneous matter-dominated expanding universe. 

Density perturbations are introduced by writing  
\begin{equation}
\Psi(\vec{r}, t) = \Psi_0(\vec{r}, t) + \Psi_1(\vec{r}, t)~~\ .
\label{pwv}
\end{equation}
The perturbation in the wavefunction is Fourier transformed in 
terms of co-moving wavevectors $\vec{k}$ as follows:
\begin{equation}
\Psi_1(\vec{r}, t) = \Psi_0(\vec{r}, t) \int d^3k~
\Psi_1(\vec{k}, t) e^{i {\vec{k}\cdot\vec{r} \over a(t)}}~~\ .
\label{FT}
\end{equation}
The Schr\"odinger-Poisson equations are satisfied to linear order
provided
\begin{equation}
\Psi_1(\vec{k}, t) = {1 \over 2} \delta(\vec{k}, t) 
+ i {m a(t)^2 \over k^2} \partial_t \delta(\vec{k}, t) 
\label{linsol}
\end{equation}
and
\begin{equation}
\partial_t^2 \delta(\vec{k},t) +
2 H(t) \partial_t \delta(\vec{k},t)
-4\pi G\rho \delta(\vec{k},t)+\frac{k^4}{4m^2a^4(t)}\delta(\vec{k},t)=0~~\ .
\label{denev}
\end{equation}
The $\delta(\vec{k}, t)$ are the Fourier components of the 
density contrast
\begin{equation}
\delta(\vec{r},t)=\frac{n_1(\vec{r},t)}{n_0(\vec{r},t)}
=\int d^3k\,\,\delta(\vec{k},t){\rm e}^{i\frac{\vec{k}\cdot\vec{r}}{a(t)}}~~\ ,
\label{dencon}
\end{equation}
where $n_1(\vec{r}, t)$ is the density perturbation.  The Fourier components
of the velocity perturbation are given by 
\begin{equation}
\vec{v}_1(\vec{k},t)=\frac{ia(t)\vec{k}}{\vec{k}\cdot\vec{k}}
\partial_t\delta(\vec{k},t)~~\ .
\label{velper}
\end{equation}
Eqs.~(\ref{denev}) - (\ref{velper}) are the standard equations 
describing the evolution of density perturbations in an expanding 
matter dominated universe except for the last term of Eq.~(\ref{denev}).
That term is absent if the matter is non-degenerate cold collisionless 
particles, such as WIMPs or sterile neutrinos.  It is due to the effect 
of the ``quantum pressure" $q$ in Eq.~(\ref{Euler}) when the matter 
is a wave.  It implies a Jeans length \cite{Khlopov,Bianchi,CABEC,Chav}
\begin{equation}
\ell_J = (16 \pi G \rho m^2)^{-{1 \over 4}} =
1.01 \cdot 10^{14} {\rm cm}
\left({10^{-5} {\rm eV} \over m}\right)^{1 \over 2}
\left({10^{-29} {\rm g/cm}^3 \over \rho}\right)^{1 \over 4}~~\ .
\label{Jeans}
\end{equation}
For $k > {a(t) \over \ell_J} = k_J$,  the density perturbations oscillate 
in time.  For $k << {a(t) \over \ell_J}$, the most general solution 
of Eq.~(\ref{denev}) is
\begin{equation}
\delta(\vec{k},t)=A(\vec{k})\left(\frac{t}{t_0}\right)^{2 \over 3}+
B(\vec{k})\left(\frac{t_0}{t}\right)~~\ ,
\label{stand}
\end{equation}
in the critical universe case ($K= 0$) where $a(t) \propto t^{2 \over 3}$.

Before we discuss the quantum evolution of the initially homogeneous 
expanding universe, let us point out that the wavefunction $\Psi_0$ in 
Eq.~(\ref{homun}) satisfies Eq.~(\ref{eom11}) with the gravitational 
potential 
\begin{equation}
\Phi_0(r,t) = {2 \pi \over 3} G m n_0(t) r^2
\label{0gp}
\end{equation}
which is indeed an appropriate solution of the Poisson equation,
Eq.~(\ref{Poiss}), but which differs from Eq.~(\ref{Newpot}) by a 
constant that diverges in the infinite volume limit.  The classical 
description above uses the Schr\"odinger-Poisson equations,  
Eqs.~(\ref{eom11}) and (\ref{Poiss}).  However, to obtain the quantum 
evolution, we will find it more convenient to start with a solution
of Eqs.~(\ref{eom11}) and (\ref{Newpot}).  The wavefunction and 
gravitational potential that describe the homogeneous expanding 
universe and solve Eqs.(\ref{eom11}) and (\ref{Newpot}) are 
\begin{eqnarray}
\Psi_0(\vec{r}, t) &=&\sqrt{n_0(t)} 
e^{i {1 \over 2} m H(t) r^2 
- i m \int^t dt^\prime~ \Phi_0(0,t^\prime)}
\nonumber\\
\Phi_0(r,t) &=& {2 \pi \over 3} G m n_0(t) r^2 + \Phi_0(0,t)~~\ .
\label{homun2}
\end{eqnarray}
The wavefunction given in Eq.~(\ref{homun2}) is the starting point 
for our discussion in the next subsection.

\subsection{Quantum evolution}

In this subsection, we derive the quantum evolution of a universe that 
starts off being described by the homogeneous expanding universe solution 
$\Psi_0$ of the classical equations of motion, Eqs.~(\ref{homun2}).  Again 
we are interested to see how long the classical solution gives a description 
consistent with quantum evolution.  We use the general method presented in 
subsection III.B. 

For a general solution $\Psi(\vec{r}, t)$ of the classical field
equation (\ref{eom11}), we expand the quantum $\psi(\vec{r}, t)$ as 
we did for $\lambda \phi^4$ theory, Eqs.~(\ref{Pmodes}) - (\ref{exp}).  
The interaction coefficients are
\begin{eqnarray}
\Lambda_{\vec{k}_1 \vec{k}_2}^{\vec{k}_3 \vec{k}_4} (t) = 
- {G m^2 \over N^2} \int_V d^3r \int_V d^3r^\prime~
{n(\vec{r}, t) n(\vec{r}^{\,\prime}, t) \over |\vec{r} - \vec{r}^{\,\prime}|}
[e^{i(\vec{k_3}-\vec{k}_1)\cdot\vec{\chi}(\vec{r},t) +
i(\vec{k}_4 - \vec{k}_2)\cdot\vec{\chi}(\vec{r}^{\,\prime},t)}  
\nonumber\\
+ e^{i(\vec{k_4}-\vec{k}_1)\cdot\vec{\chi}(\vec{r},t) +
i(\vec{k}_3 - \vec{k}_2)\cdot\vec{\chi}(\vec{r}^{\,\prime},t)}]~~\ ,
\label{gLamg}
\end{eqnarray}
and the kinetic coefficients are
\begin{eqnarray}
M_{\vec{k}}^{\vec{k}^\prime}(t) &=&
- {m \over N} \int_V d^3r~\Phi(\vec{r}, t) n(\vec{r}, t) 
e^{i(\vec{k}^\prime - \vec{k})\cdot\vec{\chi}(\vec{r}, t)}
\nonumber\\
&+& {1 \over 2 m N} \int_V d^3r~n(\vec{r}, t) 
\vec{\nabla}(\vec{k}\cdot\vec{\chi}(\vec{r}, t)) \cdot
\vec{\nabla}(\vec{k}^\prime\cdot\vec{\chi}(\vec{r}, t))
e^{i(\vec{k}^\prime - \vec{k})\cdot\vec{\chi}(\vec{r}, t)}~~\ . 
\label{gMg}
\end{eqnarray}
Eq.~(\ref{gMg}) is obtained by following the same steps as in  
the Appendix but for the gravitational case.  Note that the 
self-consistency condition Eq.~(\ref{const}) is always satisfied.

For the special solution $\Psi_0(\vec{r}, t)$ describing a 
homogeneous expanding universe, $\psi(\vec{r}, t)$ is expanded
into the orthonormal complete set of wavefuctions 
\begin{equation}
u^{\vec{k}}(\vec{r}, t) = {1 \over \sqrt{N}} 
\Psi_0(\vec{r}, t) e^{i {\vec{k}\cdot\vec{r} \over a(t)}}
= \sqrt{n_0(t) \over N} e^{i m H(t) r^2 - i m \int^t dt^\prime \Phi_0(0,t^\prime) 
+ i {\vec{k}\cdot\vec{r} \over a(t)}}~~\ .
\label{ONC3}
\end{equation}
The $u^{\vec{k}}(\vec{r}, t)$ are similar to $\Psi_0$ but 
differ from it by long wavelength modulations.  They have the 
properties described by Eqs.~(\ref{Pmodes}) - (\ref{ONC2}), with 
$\vec{\chi}(\vec{r}, t) = {\vec{r} \over a(t)}$.  We specialize 
henceforth to the critical universe ($K = 0$) for which 
\begin{equation}
a(t) = \left({t \over t_*}\right)^{2 \over 3}~~~{\rm and}~~~
n(t) = n_* \left({t_* \over t}\right)^2
\label{crit}
\end{equation}
where $t_*$ is an arbitrarily chosen initial time.  The 
interaction coefficients in the basis of Eq.~(\ref{ONC3}) 
are in that case
\begin{equation}
\Lambda_{\vec{k}_1 \vec{k}_2}^{\vec{k}_3 \vec{k}_4}(t) = 
- {4 \pi G m^2 \over V_*} \left({t_* \over t}\right)^{2 \over 3} 
\delta_{\vec{k}_1 + \vec{k}_2}^{\vec{k}_3 + \vec{k}_4}
\left({1 \over (\vec{k}_4 - \vec{k}_1)^2 + \mu^2} + 
{1 \over (\vec{k}_3 - \vec{k}_1)^2 + \mu^2}\right)
\label{gLam2}
\end{equation}
where $V_* = N/n_*$ is the volume occupied by the system at the 
initial time $t_*$, and $\mu$ is an infrared cutoff.  The kinetic 
coefficients are 
\begin{equation}
M_{\vec{k}}^{\vec{k}^\prime}(t) = 
\left({\vec{k}\cdot\vec{k}^\prime \over 2 m} 
\left({t_* \over t}\right)^{4 \over 3}
+ {2 m \over 3 t_*^2} \left({t_* \over t}\right)^{2 \over 3} 
{1 \over (\vec{k} - \vec{k}^\prime)^2 + \mu^2}\right)
\delta_{\vec{k}}^{\vec{k}^\prime}~~\ .
\label{gM}
\end{equation}
The equations of motions for the $a_{\vec{k}}(t)$ operators 
and their classical analogs $A_{\vec{k}}$ are the same as in 
the previous section, Eqs.~(\ref{eom3}) and (\ref{clm3}), but 
with the $\Lambda_{\vec{k}_1 \vec{k}_2}^{\vec{k}_3 \vec{k}_4}$
and $M_{\vec{k}}^{\vec{k}^\prime}$ given by the above expressions.  
The consistency condition Eq.~(\ref{const}) is satisfied since 
the Friedmann equation implies 
\begin{equation}
4 \pi G m n_* = {2 \over 3 t_*^2}~~\ .
\label{FR5}
\end{equation}
The consistency condition ensures that 
\begin{equation}
A_{\vec{k}}(t) = \sqrt{N} \delta_{\vec{k}}^{\vec{0}}~~\ ,
\label{clsol2}
\end{equation}
which describes the homogeneous expanding universe, is a solution of 
the classical equations of motion.  

To analyze the behavior of the quantum system when the homogeneous 
particle state ($\vec{k} = 0$) is occupied by a huge number $N$ of 
quanta, we substitute
\begin{equation}
a_{\vec{k}}(t) = \sqrt{N} \delta_{\vec{k}}^{\vec{0}} + b_{\vec{k}}(t)~~\ . 
\label{shift}
\end{equation}
The $b_{\vec{k}}(t)$ satisfy canonical commutation relations, and 
the equations of motion
\begin{eqnarray}
i \partial_t b_{\vec{k}} &=& (M_{\vec{k}}^{\vec{k}^\prime} + 
N \Lambda_{\vec{k}\,\vec{0}}^{\vec{k}^\prime \vec{0}}) b_{\vec{k}^\prime}
+ {1 \over 2} N \Lambda_{\vec{k} \vec{k}^\prime}^{\vec{0} \vec{0}}
b_{\vec{k}^\prime}^\dagger + ... \nonumber\\
&=& \left({\vec{k}\cdot\vec{k} \over 2 m} \left({t_* \over t}\right)^{4 \over 3} 
- {2 m \over 3 t_*^2} \left({t_* \over t}\right)^{2 \over 3} 
{1 \over \vec{k}\cdot\vec{k}}\right) b_{\vec{k}}
- {2 m \over 3 t_*^2} \left({t_* \over t}\right)^{2 \over 3} 
{1 \over \vec{k}\cdot\vec{k}}~b_{-\vec{k}}^\dagger + ...
\label{geom}
\end{eqnarray}
where the dots represent interaction terms.  The interaction terms
are suppressed by one or two factors of $1/\sqrt{N}$ and will be 
ignored henceforth.

\subsection{Instability by parametric resonance}

Eq.~(\ref{geom}) may be rewritten
\begin{equation}
i \partial_t \left( \begin{array} {c} b_{\vec{k}}
\\ b_{-\vec{k}}^\dagger \end{array}\right) =
\left( \begin{array} {cc} A(t)~~ & ~~B(t) \\
- B(t) ~~ &~~ - A(t) \end{array} \right)
\left( \begin{array} {c} b_{\vec{k}}
\\ b_{-\vec{k}}^\dagger \end{array}\right) ~~\ , 
\label{ignint2}
\end{equation}
where $A(t) \equiv \epsilon(t) - \gamma(t)$, $B(t) = - \gamma(t)$, 
and  
\begin{equation}
\epsilon(t) = {k^2 \over 2 m} \left({t_* \over t}\right)^{4 \over 3} 
~~~ ,~~~
\gamma(t) = {1 \over 3 t_*^2} \left({t_* \over t}\right)^{2 \over 3} 
{2 m \over k^2}~~\ .
\label{ABdef}
\end{equation}
We perform a time-dependent Bogoliubov transformation
\begin{equation}
\left( \begin{array} {c} b_{\vec{k}}
\\ b_{-\vec{k}}^\dagger \end{array}\right) =
\left( \begin{array} {cc} c(t)~~ & ~~s(t) \\                                 
s(t)^* ~~ &~~ c(t)^* \end{array} \right)                                 
\left( \begin{array} {c} \beta_{\vec{k}}
\\ \beta_{-\vec{k}}^\dagger \end{array}\right) ~~\ .
\label{Bog2}
\end{equation}
The transformation is canonical provided 
$|c(t)|^2 - |s(t)|^2 = 1$, and provided $c(t)$ and 
$s(t)$ do not depend on the sign of $\vec{k}$.  (The $\vec{k}$
dependence of $A$, $B$, $\epsilon$, $\gamma$, $c$, $s$ is 
suppressed to avoid cluttering the equations unnecessarily.)
The new operators satisfy 
\begin{equation}
i \partial_t \left( \begin{array} {c} \beta_{\vec{k}}
\\ \beta_{-\vec{k}}^\dagger \end{array}\right) =
\left( \begin{array} {cc} {\cal A}(t)~~ & ~~{\cal B}(t) \\                                 
- {\cal B}(t)^* ~~ &~~ - {\cal A}(t)^* \end{array} \right)                                 
\left( \begin{array} {c} \beta_{\vec{k}}
\\ \beta_{-\vec{k}}^\dagger \end{array}\right) ~~\ ,
\label{eom33}
\end{equation}
where 
\begin{eqnarray}
{\cal A}(t) &=&(|c|^2 + |s|^2) A + (cs + c^*s^*) B 
- i(c^* \dot{c} - s \dot{s}^*)\nonumber\\
{\cal B}(t) &=& 2 c^* s A + (c^{*2} + s^2) B
- i(c^* \dot{s} - s \dot{c}^*)~~\ .
\label{calAB}
\end{eqnarray}
The Jeans length, Eq.~(\ref{Jeans}), increases as $t^{1\over 2}$, 
whereas the wavelength associated with each wavevector $\vec{k}$ 
increases as $a(t) \propto t^{2 \over 3}$.  Hence there is for 
each wavevector $\vec{k}$ a time of order
\begin{equation}
t_k = {k^6 t_*^4 \over (2 m)^3}
\label{tk}
\end{equation}
before which the perturbations with that wavevector are stable and 
after which they are unstable.   

Consider modes that are deeply in the unstable regime at the time 
$t$ under consideration, i.e. $t_k << t$.  These are the modes that 
obey Eq.~(\ref{stand}) in the classical description.  We may set 
$\cal{A}$ = 0 by choosing $c = \cosh(u)$, $s = \sinh(u)$ with 
\begin{equation}
\tanh(2u) = - {A \over B} = -1 + {\epsilon \over \gamma}~~\ .
\label{th2u}
\end{equation}
Since $\epsilon << \gamma$, $u$ is large and negative.   We have to 
leading order 
\begin{equation}
u = {1 \over 4} \ln\left({\epsilon \over 2 \gamma}\right) = {1 \over 4} 
\ln\left[{3 \over 2}\left({t_k \over t}\right)^{2 \over 3}\right]~~\ , 
\label{eqfu}
\end{equation}
and therefore
\begin{equation}
{\cal B} = - \sqrt{B^2 - A^2} - i \dot{u} = 
- \sqrt{4 \pi G m n_0(t)} + {i \over 6 t}~~\ .
\label{eqfB}
\end{equation}
The equation of motion for the $\beta_{\vec{k}}$ operators is thus
\begin{equation}
i \partial_t \beta_{\vec{k}} = 
( - \sqrt{2 \over 3} + {i \over 6}){1 \over t} \beta_{-\vec{k}}^\dagger~~\ .
\label{eom25}
\end{equation}
The Hamiltonian for the modes of wavevector $\vec{k}$ and $-\vec{k}$,
with $t_k << t$, is thus
\begin{equation}
H_{\vec{k}} = ( - \sqrt{2 \over 3} + {i \over 6}){1 \over t} 
\beta_{-\vec{k}}^\dagger \beta_{\vec{k}}^\dagger + 
( - \sqrt{2 \over 3} - {i \over 6}){1 \over t} 
\beta_{-\vec{k}} \beta_{\vec{k}}~~\ .
\label{Ham25}
\end{equation}
It may be rewritten
\begin{eqnarray}
H_{\vec{k}} &=& - {1 \over 2 t}
(\eta \alpha_{\vec{k}} \alpha_{\vec{k}} + 
\eta^* \alpha_{\vec{k}}^\dagger \alpha_{\vec{k}}^\dagger) 
+ {1 \over 2 t}(\eta \alpha_{\vec{k}}^\prime \alpha_{\vec{k}}^\prime + 
\eta^* \alpha_{\vec{k}}^{\prime\dagger} \alpha_{\vec{k}}^{\prime\dagger})
\label{Ham25p}
\end{eqnarray}
in terms of the canonical variables
\begin{eqnarray}
\alpha_{\vec{k}} &=& {1 \over \sqrt{2}}(\beta_{\vec{k}} + \beta_{-\vec{k}})
\nonumber\\
\alpha_{\vec{k}}^\prime &=& {1 \over \sqrt{2}}(\beta_{\vec{k}} - \beta_{-\vec{k}})
\label{canon}
\end{eqnarray}
and the constant $\eta \equiv \sqrt{2 \over 3} + {i \over 6} = |\eta|e^{i \theta}$ 
with $|\eta| = {5 \over 6}$ and $\sin\theta = {1 \over 5}$.  The phase of 
$\pm \eta$ can be absorbed into a redefinition of the $\alpha_{\vec{k}}$ and 
$\alpha_{\vec{k}}^\prime$ operators.  

We thus consider the dynamics implied by a Hamiltonian of the form
\begin{equation}
H(t) = {\xi \over 2t} (\alpha(t) \alpha(t)
+ \alpha(t)^\dagger \alpha(t)^\dagger)
\label{Ham33}
\end{equation}
where $\xi$ is a real positive constant.  The equation of motion
\begin{equation}
i \partial_t \alpha(t) = {\xi \over t} \alpha(t)^\dagger
\label{eom34}
\end{equation} 
is solved by 
\begin{equation}
\alpha(t) = 
{1 \over 2} (\alpha(t_*) -i \alpha(t_*)^\dagger) 
\left({t \over t_*}\right)^\xi 
+ {1 \over 2} (\alpha(t_*) +i \alpha(t_*)^\dagger)
\left({t_* \over t}\right)^\xi~~\ .
\label{sol34}
\end{equation}
Eq.~(\ref{sol34}) implies an instability, albeit only a power law
instability.  

To see its implications, consider the evolution of states in the 
Schr\"odinger picture.  The Schr\"odinger picture Hamiltonian is
\begin{equation}
H_s(t) = {\xi \over 2t}(\alpha(t_*)\alpha(t_*) 
+ \alpha(t_*)^\dagger \alpha(t_*)^\dagger)~~\ ,
\label{sham}
\end{equation}
and the time evolution operator
\begin{equation}
U(t,t_*) = \exp[-i {\xi \over 2} (\alpha(t_*)\alpha(t_*) +
\alpha(t_*)^\dagger \alpha(t_*)^\dagger) \ln\left({t \over t_*}\right)]~~\ .
\label{evop}
\end{equation}
We have
\begin{equation}
\alpha(t) = U(t,t_*)^\dagger \alpha(t_*) U(t, t_*)~~\ .
\label{rel}
\end{equation}
As an example, consider the evolution 
\begin{equation}
|\Psi_s(t)> = U(t, t_*) |\Psi(t_*)>
\label{Sevol}
\end{equation}
of the state defined by 
\begin{equation}
\alpha(t_*) |\Psi(t_*)> = 0~~\ .
\label{annih}
\end{equation}
Combining Eqs.~(\ref{rel})-(\ref{annih}) and (\ref{sol34}), we have
\begin{eqnarray}
U(t,t_*) \alpha(t_*) U(t,t_*)^\dagger |\Psi_s(t)> &=& \nonumber\\ 
\big[{1 \over 2} (\alpha(t_*) -i 
\alpha(t_*)^\dagger)
\left({t_* \over t}\right)^\xi
&+& {1 \over 2} (\alpha(t_*) +i \alpha(t_*)^\dagger)
\left({t\over t_*}\right)^\xi \big] |\Psi_s(t)> = 0~~\ .
\label{ann2}
\end{eqnarray}
Eq.~(\ref{ann2}) yields a recursion relation between the 
coefficients in the expansion
\begin{equation}
|\Psi_s(t)> = \sum_{n=0}^\infty c_n(t) |n>~~\ ,
\label{gexp}
\end{equation}
where
\begin{equation}
|n> = {1 \over \sqrt{n!}} (\alpha(t_*)^\dagger)^n |\Psi(t_*)>~~\ .
\label{gdefn}
\end{equation}
The recursion relation implies
\begin{eqnarray}
c_n(t) &=& 0~~~~~~~~~~~~~~~~~~~~~~~~~~~~~~~~~~~~~~~~~~~~~~{\rm for}~{n}~{\rm odd}
\nonumber\\
&=& \left(- i \tanh(s)\right)^p
\sqrt{(2p-1)!! \over 2^p~p!} c_0(t)~~~~~~~~{\rm for}~n=2p~~~\ ,
\label{c2p}
\end{eqnarray}
where $s$ is defined by 
\begin{equation}
e^{-s} = \left({t_* \over t}\right)^\xi~~\ .
\label{gedfs}
\end{equation} 
The normalization condition $<\Psi_s(t)|\Psi_s(t)> = 1$ yields then
\begin{equation}
|c_0(t)|^2 = {1 \over \cosh(s)}~~\ .
\label{grstper}
\end{equation}
The average occupation number and average occupation number squared are
\begin{eqnarray}
\langle N(t) \rangle &=& 
\langle \Psi(t_*)|\alpha(t)^\dagger \alpha(t) |\Psi(t_*)\rangle 
= \sinh^2(s) \nonumber\\
\langle N(t)^2 \rangle &=& \langle \Psi(t_*)| \alpha(t)^\dagger \alpha(t)
\alpha(t)^\dagger \alpha(t)| \Psi(t_*)\rangle =
\sinh^4(s) + 2 \sinh^2(s) \cosh^2(s)~\ .
\label{NN2}
\end{eqnarray}
The root mean square deviation of the occupation number from its
average is thus
\begin{equation}
\delta N(t) = {1 \over \sqrt{2}} \sinh(2s)~~\ .
\label{grmsd}
\end{equation}
Both the average occupation number and its root mean square 
deviation increase as $({t \over t_*})^{2\xi} = 
({t \over t_*})^{5 \over 3}$ for $\xi = |\eta| = {5 \over 3}$.

\subsection{Duration of classicality}

To lowest order in the perturbations the density operator is
\begin{eqnarray}
n(\vec{r}, t) &=& \psi(\vec{r}, t)^\dagger \psi(\vec{r}, t) \
= \sum_{\vec{k}, \vec{k}^\prime} 
u^{\vec{k}}(\vec{r}, t)^* a_{\vec{k}}(t)^\dagger
u^{\vec{k}^\prime}(\vec{r}, t) a_{\vec{k}^\prime}(t) \nonumber\\
&=& N u^{\vec{0}}(\vec{r}, t)^* u^{\vec{0}}(\vec{r}, t) +
\sqrt{N} \sum_{\vec{k} \neq 0} 
[u^{\vec{0}}(\vec{r}, t)^* u^{\vec{k}}(\vec{r} ,t) b_{\vec{k}}(t)
+ u^{\vec{0}}(\vec{r}, t) u^{\vec{k}}(\vec{r} ,t)^* b_{\vec{k}}(t)^\dagger]
+ {\cal O}(1/N)
\nonumber\\
&=& n_0(t) + {n_0(t) \over \sqrt{N}} \sum_{\vec{k} \neq 0}
[b_{\vec{k}}(t) + b_{-\vec{k}}^\dagger(t)] 
e^{i {\vec{k}\cdot\vec{r} \over a(t)}}
+ {\cal O}(1/N)~~\ .
\label{fgden}
\end{eqnarray}
Since
\begin{eqnarray}
b_{\vec{k}}(t) + b_{-\vec{k}}(t)^\dagger
&=&(c(t)+s(t))(\beta_{\vec{k}}(t) + \beta_{-\vec{k}}(t)^\dagger)
\nonumber\\
&=& e^u {1 \over \sqrt{2}} (\alpha_{\vec{k}}(t)
+ \alpha^\prime_{\vec{k}}(t) + 
\alpha_{\vec{k}}(t)^\dagger - \alpha_{\vec{k}}^\prime(t)^\dagger)
\nonumber\\ 
&\propto& t^{-{1 \over 6}}(t^{5 \over 6}~~{\rm and}~~t^{-{5 \over 6}})
= t^{2 \over 3}~~{\rm and}~~t^{-1}~~\ ,
\label{see}
\end{eqnarray}
we see that the perturbations grow at the same rate as in the 
classical description, Eq.~(\ref{stand}).  The main difference 
is that the perturbations are seeded in the quantum description 
whereas in the classical description they are not.

The $b_{\vec{k}}(t)$ annihilation operators (for $\vec{k} \neq 0$)
are given in terms of the $\alpha_{\vec{k}}(t_*)$ and  
$\alpha_{\vec{k}}^{\prime}(t_*)$ by 
\begin{eqnarray}
b_{\vec{k}}(t) &=& 
{c(t) \over \sqrt{2}} (\alpha_{\vec{k}}(t) + \alpha^\prime_{\vec{k}}(t))
+ {s(t) \over \sqrt{2}} 
(\alpha_{\vec{k}}^\dagger(t) - \alpha_{\vec{k}}^{\prime~\dagger}(t))
\nonumber\\ 
&=& 
{1 \over 2 \sqrt{2}} \left({t \over t_*}\right)^{5 \over 6}
[(c(t) - i e^{i \theta} s(t))
\left(\alpha_{\vec{k}}(t_*) + \alpha^\prime_{\vec{k}}(t_*)\right)
\nonumber\\
&+& (s(t) + i e^{- i \theta} c(t)) \left(\alpha_{\vec{k}}^\dagger(t_*) 
- \alpha_{\vec{k}}^{\prime~\dagger}(t_*)\right)]
\label{bridge}
\end{eqnarray}
where we used Eqs.~(\ref{Bog2}) and (\ref{canon}), and Eq.~(\ref{sol34}) 
with $\xi = |\eta| = {5 \over 6}$ for $\alpha(t) = e^{i \theta/2} 
\alpha^\prime_{\vec{k}}(t)$
and $\alpha(t) = i e^{i \theta/2} \alpha_{\vec{k}}(t)$.  We kept growing 
terms only.  For $t >> t_k$ we have
\begin{equation}
c(t) = - s(t) = {1 \over 2} e^{- u} = 
{1 \over 2} \left({2 \over 3}\right)^{1 \over 4} 
\left({t \over t_k}\right)^{1 \over 6}
\label{subs}
\end{equation}
in view of Eq.~(\ref{eqfu}).  Therefore
\begin{equation}
b_{\vec{k}}(t) = {1 \over 4 \sqrt{2}} \left({2 \over 3}\right)^{1 \over 4}
\left({t \over t_*}\right)^{5 \over 6} \left({t \over t_k}\right)^{1 \over 6}
[(1 + i e^{i \theta})(\alpha_{\vec{k}}(t_*) + \alpha^\prime_{\vec{k}}(t_*))
- (1 - i e^{- i \theta})(\alpha_{\vec{k}}^\dagger(t_*) - 
\alpha_{\vec{k}}^{\prime~\dagger}(t_*))]
\label{phd}
\end{equation}
for $t_k < t_* << t$.  For $t_* < t_k << t$ we replace $t_*$ by $t_k$ in the 
above expression since a mode starts to grow only at time $t_k$.

The discussion in the previous subsection suggests that the state 
with the longest duration of classicality for describing a homogeneous
condensate is $|\Psi \rangle$:
\begin{equation}
\alpha_{\vec{k}}(t_*)|\Psi\rangle = 
\alpha^\prime_{\vec{k}}(t_*)|\Psi \rangle = 0~~\ .
\label{gstate}
\end{equation}
In view of Eq.~(\ref{phd}) and the sentence following, the average number 
of quanta that have jumped from the condensate into mode $b_{\vec{k}}$  
with physical wavevector magnitude ${k \over a(t)} < \ell_J(t)^{-1}$ is 
\begin{eqnarray}
\langle N_{\vec{k}}(t)\rangle = 
\langle \Psi| b_{\vec{k}}^\dagger(t) b_{\vec{k}}(t) |\Psi>
&\simeq& {1 \over 10} \sqrt{2 \over 3} \left({t \over t_*}\right)^2
\left({t_* \over t_k}\right)^{1 \over 3}~~~{\rm for}~t_k < t_* << t
\nonumber\\
&\simeq& {1 \over 10} \sqrt{2 \over 3} \left({t \over t_k}\right)^2
~~~~~~~~~~~~~{\rm for}~t_* < t_k << t
\label{gNkt}
\end{eqnarray}
in state $|\Psi \rangle$ at time $t$.  We used $\sin\theta = {1 \over 5}$.  
As an alternative to $|\Psi \rangle$ we considered the state 
$|\Psi^\prime \rangle$ defined by 
\begin{equation}
b_{\vec{k}}(t_*)|\Psi^\prime \rangle = 0
\label{alt}
\end{equation}
for all $\vec{k} \neq 0$, and verified that the average occupation number 
for any $\vec{k} \neq 0$ mode is larger in state $|\Psi^\prime \rangle$ 
than in state $|\Psi \rangle$ for large $t$.  The state $|\Psi \rangle$ 
thus has the larger duration of classicality and is the state that we 
consider henceforth.

In state $|\Psi \rangle$ the total number of quanta that have left 
the $\vec{k} = 0$ condensate at time $t$ is  
\begin{eqnarray}
N_{\rm ev}(t) &=& \sum_{k < k_J} \langle N_{\vec{k}}(t) \rangle
\nonumber\\ 
&\sim& V_* {1 \over 10} \sqrt{2 \over 3} \left({t \over t_*}\right)^2
[\int_{k < k_J(t_*)} {d^3 k \over (2 \pi)^3} {2m \over k^2 t_*} 
+ \int_{k > k_J(t_*)} {d^3 k \over (2 \pi)^3}
\left({2m \over k^2 t_*}\right)^6]
\nonumber\\
&\sim& 1.3 {V_* k_J(t_*)^3 \over 20 \pi^2} \left({t \over t_*}\right)^2
\sim 0.26 N  G m^2 \sqrt{m t_*} 
\left({t \over t_*}\right)^2~~\ ,
\label{gNp}
\end{eqnarray}
where we used Eqs.~(\ref{Jeans}), (\ref{FR5}) and (\ref{tk}).   
The integral over $\vec{k}$ in Eq.~(\ref{gNp}) should be restricted
to $k > a(t) H(t)$ since the modes are unstable only for 
wavelengths that are within the horizon.  However, this 
restriction is irrelevant since the integral is dominated 
by values of $k$ near $k_J(t_*)$.  After a time of order
\begin{equation}
t_c \sim t_* {1 \over (G m^2 \sqrt{mt_*})^{1 \over 2}}
\label{gtc}
\end{equation}
the $\vec{k} = 0$ condensate is largely depleted.

\section{Summary}

This paper sought to clarify aspects of the evolution of the cosmic 
axion dark matter fluid.  It was found in refs.\cite{CABEC,axtherm}
that dark matter axions thermalize by gravitational self-interactions.
When they thermalize all conditions for their Bose-Einstein condensation
are satisfied, and we expect therefore that this is indeed what happens.
Furthermore, it was shown that axion Bose-Einstein condensation explains 
in detail and in all respects the evidence for caustic rings of dark 
matter \cite{case}.  Nonetheless, axion Bose-Einstein condensation is 
a difficult subject from the theoretical point of view.  The central 
difficulty is that gravity, the interaction by which axions thermalize, 
causes instability.  Bose-Einstein condensation means that most of the 
particles go to their lowest energy state.  But, if the system is unstable 
it is not obvious what is the lowest energy state.   Ref. \cite{Guth} 
concluded that "while a Bose-Einstein condensate is formed, the claim 
of long-range correlation is unjustified.".

Section II was written in response to the critique of 
ref. \cite{Davidson,Davidson2,Guth}.
We emphasize that Bose-Einstein condensation is a quantum phenomenon, 
even if some aspects of Bose-Einstein condensation can be reproduced 
in a truncated classical field theory.  We reiterate the conclusion of 
ref. \cite{simtherm} that the classical description necessarily differs 
from the quantum description on the thermalization time scale.  We show 
that a Bose-Einstein condensate always has long range correlations, 
whether or not it is homogeneous.  Finally we clarify that Bose-Einstein 
condensation is always into the lowest energy state available through the 
thermalizing interactions.  Remaining questions are: what is in general 
the state that the axions condense into, or move towards (since thermal 
equilibrium is always incomplete in an unstable system)?  How does one
determine it?  Or, more broadly:  what is the evolution of a degenerate 
quantum scalar field as a result of its gravitational self-interactions?

The evidence for cosmic axion Bose-Einstein condensation from caustic 
rings does not demand the clarifications that we seek in this paper.
Indeed that evidence is based on the unambiguous statement that 
the lowest energy state for given total angular momentum is a state 
of rigid rotation in the angular variables.  There is no instability 
in the angular variables.  Gravitational instability resides in the 
scalar modes, not in the rotational vector modes.

Cosmological perturbation theory is an arena where we do seek 
clarification.  Its results are consistent with a wavefunction 
description of cold dark matter.   The wavefunction $\Psi(\vec{x}, t)$
is a solution of the classical field equations.  On the other hand, 
it is only one mode of the quantum field $\psi(\vec{x}, t)$.  Given 
such a classical description, where do the quantum corrections appear 
and how large are they?  Although we were not able to answer this 
question in general, we made progress.

In Section III, we expanded the quantum scalar field in a set of modes 
built around an arbitrary solution $\Psi(\vec{x}, t)$ of the classical 
field equations; Eqs. (\ref{Pmodes})-(\ref{eom3}).  The modes are 
labeled by a wavevector $\vec{k}$ which is conjugate to the co-moving 
coordinates defined by the flow that the classical solution describes.  
The classical solution itself is mode $\vec{k} = 0$.  We derived the 
Hamiltonian in terms of the creation and annihilation operators for the 
$\vec{k}$-modes. The kinetic coefficients $M_{\vec{k}}^{\vec{k}^\prime}(t)$ and 
interaction coefficients $\Lambda_{\vec{k}_1 \vec{k}_2}^{\vec{k}_3 \vec{k}_4}(t)$ 
that appear in the Hamiltonian are functionals of the classical solution 
$\Psi(\vec{x}, t)$.  

We applied the formalism to the homogeneous condensate in $\lambda\phi^4$ 
theory.  In the repulsive case ($\lambda > 0$) our treatment merely 
reproduces well-known results.  In the attractive case ($\lambda < 0$), 
we show that the condensate becomes depleted by parametric resonance: 
pairs of quanta jump out of the condensate into each mode with wavector 
less than a critical value $k_J = \sqrt{|\lambda| n_0 \over 2 m}$, 
where $n_0$ is particle density and $m$ is particle mass.  The occupation 
number of each state with $k < k_J$ grows exponentially at the rate 
${k \over m}\sqrt{k_J^2 - k^2}$.  We calculate the time $t_c$ after 
which the condensate is almost entirely depleted, Eq.~(\ref{cdur}).  
In contrast, according to the classical equations of motion, the 
homogeneous condensate persist forever.  Our treatment of the 
quantum system is exact in the limit where the number of quanta 
in the condensate $N$ is large and time $t << t_c$.

In Section IV, we applied the formalism to a self-gravitating 
Bosonic fluid forming a homogeneous, critically expanding  
universe.  The classical solution describing the homogeneous 
expanding universe is given in Eq.~(\ref{homun}).  In the 
quantum description, parametric resonance causes pairs of 
quanta to jump out of the condensate into each mode with 
wavector $k < k_J = 1/\ell_J$ where $\ell_J$ is the Jeans length, 
Eq.~(\ref{Jeans}).  The occupation number of each mode with $k < k_J$ 
grows as a power law, as in the classical description.  As for 
$\lambda \phi^4$ theory with $\lambda < 0$, the main difference 
between the classical and quantum descriptions is that in the 
quantum description the $0 < k <k_J$ perturbations are seeded 
whereas in the classical description they are not.  Whereas the 
homogeneous condensate persists forever in the classical description,    
it is depleted after a time $t_c$ in the quantum description; $t_c$ 
is given in Eq.~(\ref{gtc}) for the self-gravitating case.  Our 
treatment of the self-gravitating quantum system is exact for all
modes with $k << k_J$, in the limit where $N$ is large and $t << t_c$.

Although we analyzed only homogeneous condensates, the fact that 
quantum evolution differs from classical evolution after a time 
must be true for inhomogeneous condensates as well since a homogeneous 
condensate is a limiting case of inhomogeneous condensates.  In fact, 
taking as a guide our analysis of the 5 oscillator model in Section II.2, 
we expect that inhomogeneous condensates have a shorter duration of 
classicality than the homogeneous condensate.  The behavior of 
inhomogeneous condensates will be addressed in future work.

We used Newtonian gravity throughout our discussion of the 
homogeneous self-gravitating condensate.  As already mentioned, 
this is valid only when the velocities are small compared to 
the speed of light.  Our treatment applies therefore only to 
modes that are well within the horizon ($k/a(t) >> H(t)$).
Before they enter the horizon, the modes are frozen by 
causality.  They do not grow then and hence do not contribute
to the depletion of a condensate.  A general relativistic 
treatment is necessary to obtain a description of a mode
as it enters the horizon and begins to contribute to 
condensate depletion.  

\begin{acknowledgments}

We are grateful to Edward Witten, Charles Thorn, Adam Christopherson 
and Nilanjan Banik for useful discussions.  PS thanks the CERN Theory 
Group for its hospitality and support.  This work was supported in part 
by the U.S. Department of Energy under grant DE-FG02-97ER41209 and by 
the Heising-Simons Foundation under grant No. 2015-109.

\end{acknowledgments}


\appendix

\section{}

This appendix provides a derivation of Eq.~(\ref{M2}).  The 
expression for $M_{\vec{k}}^{\vec{k}^\prime}(t)$ 
in Eq.~(\ref{MLam}) may be rewritten
\begin{equation}
M_{\vec{k}}^{\vec{k}^\prime} = \int_V d^3x~{1 \over 2}
[u^{\vec{k}*}(- i \partial_t - {\nabla^2 \over 2m}) u^{\vec{k}^\prime}
+ \left((- i \partial_t - {\nabla^2 \over 2m}) u^{\vec{k}}\right)^*
u^{\vec{k}^\prime}]
\label{A1}
\end{equation}
by integrating by parts and noting that 
\begin{equation}
\int_V d^3x~[u^{\vec{k}*} \partial_t u^{\vec{k}^\prime}
+ \partial_t u^{\vec{k}*}~u^{\vec{k}^\prime}] = 0
\label{A2}
\end{equation}
in view of Eq.~(\ref{ONC2}).  Substituting Eq.~(\ref{Pmodes}) and 
using the fact that $\Psi(\vec{x}, t)$ satisfies Eq.~(\ref{SGP})
one obtains:
\begin{eqnarray}
M_{\vec{k}}^{\vec{k}^\prime} &=& {1 \over 2 N} \int_V d^3x~
e^{i (\vec{k}^\prime - \vec{k})\cdot\vec{\chi}(\vec{x}, t)}
\Big[- {\lambda \over 4 m^2} |\Psi(\vec{x}, t)|^4
\nonumber\\
&+& |\Psi(\vec{x}, t)|^2 \Big(
(\vec{k}^\prime + \vec{k})\cdot\partial_t\vec{\chi}
- {i \over m}\vec{\nabla}\ln\Psi\cdot \vec{\nabla}(\vec{k}^\prime\cdot\vec{\chi})
+ {i \over m} \vec{\nabla} \ln \Psi^* \cdot \vec{\nabla} (\vec{k}\cdot\vec{\chi})
\nonumber\\
&+& {i \over 2m} \nabla^2 \left((\vec{k} - \vec{k}^\prime)\cdot\vec{\chi}\right)
+ {1 \over 2m} \left(\vec{\nabla}(\vec{k}\cdot\vec{\chi})\right)^2
+ {1 \over 2m} \left(\vec{\nabla}(\vec{k}^\prime\cdot\vec{\chi})\right)^2\Big)
\Big]~~\ .
\label{long}
\end{eqnarray}
Since
\begin{equation}
\vec{\nabla} \ln \Psi = {1 \over 2} \vec{\nabla} \ln (n) 
+ i m \vec{v}
\label{nmv}
\end{equation}
we have 
\begin{eqnarray}
{i \over m}[\vec{\nabla}\ln\Psi^*\cdot \vec{\nabla}(\vec{k}\cdot\vec{\chi})
&-& \vec{\nabla}\ln\Psi \cdot \vec{\nabla} (\vec{k}^\prime\cdot\vec{\chi})]
\nonumber\\
= {i \over 2m} \vec{\nabla} \ln (n) \cdot 
\vec{\nabla} \left((\vec{k} - \vec{k}^\prime)\cdot\vec{\chi}\right) 
&+& \vec{v}\cdot \vec{\nabla} \left((\vec{k} 
+ \vec{k}^\prime)\cdot\vec{\chi}\right)~~\ .
\label{inter}
\end{eqnarray}
Because
\begin{equation}
(\partial_t + \vec{v}\cdot\vec{\nabla}) \vec{\chi} = 0~~\ , 
\label{folmot}
\end{equation}
Eq.~(\ref{long}) simplifies to
\begin{eqnarray}
M_{\vec{k}}^{\vec{k}^\prime} &=& 
- {\lambda \over 8 m^2} \tilde{n}(\vec{k} - \vec{k}^\prime,t) 
+ {1 \over 2 N} \int_V d^3x~n(\vec{x}, t) 
e^{i(\vec{k}^\prime - \vec{k})\cdot\vec{\chi}(\vec{x}, t)}
\Big[{1 \over 2m} \left(\vec{\nabla}(\vec{k}\cdot\vec{\chi})\right)^2 
\nonumber\\ 
&+& {1 \over 2m} \left(\vec{\nabla}(\vec{k}^\prime\cdot\vec{\chi})\right)^2
+ {i \over 2m}\vec{\nabla}\ln (n)\cdot
\vec{\nabla}\left((\vec{k}-\vec{k}^\prime)\cdot\vec{\chi}\right)
+ {i \over 2m} \nabla^2 \left((\vec{k} - \vec{k}^\prime)\cdot\vec{\chi}\right)\Big] 
~~\ .
\label{shorter}
\end{eqnarray}
Upon integrating the third term in brackets by parts, one finds Eq.~(\ref{M2}).


\newpage



\newpage


\begin{figure}
\begin{center}
\includegraphics[height=150mm]{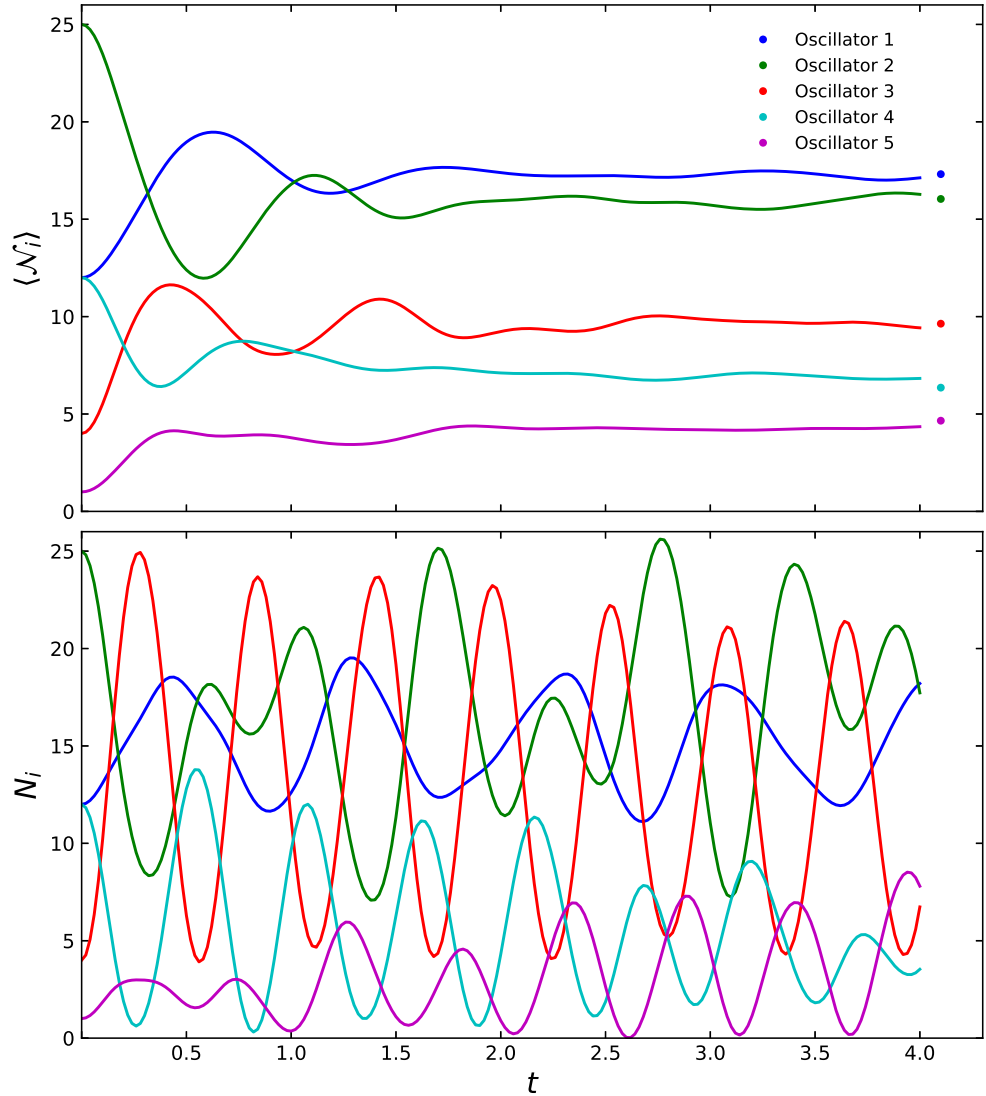}
\vspace{0.3in}
\caption{Quantum (top) and classical (bottom) time evolution         
of the occupation numbers in the toy system described in the         
text for the initial state $|12,25,4,12,1\rangle$.  The dots        
on the right in the top panel indicate the thermal averages          
in the quantum case.  The quantum system approaches the thermal 
averages on the expected time scale.   The classical evolution 
tracks the quantum evolution only very briefly and does not 
equilibrate.  This figure is taken from ref. \cite{simtherm}.}
\end{center}
\label{quantclass}
\end{figure}

\newpage

\begin{figure}
\begin{center}
\includegraphics[height=150mm]{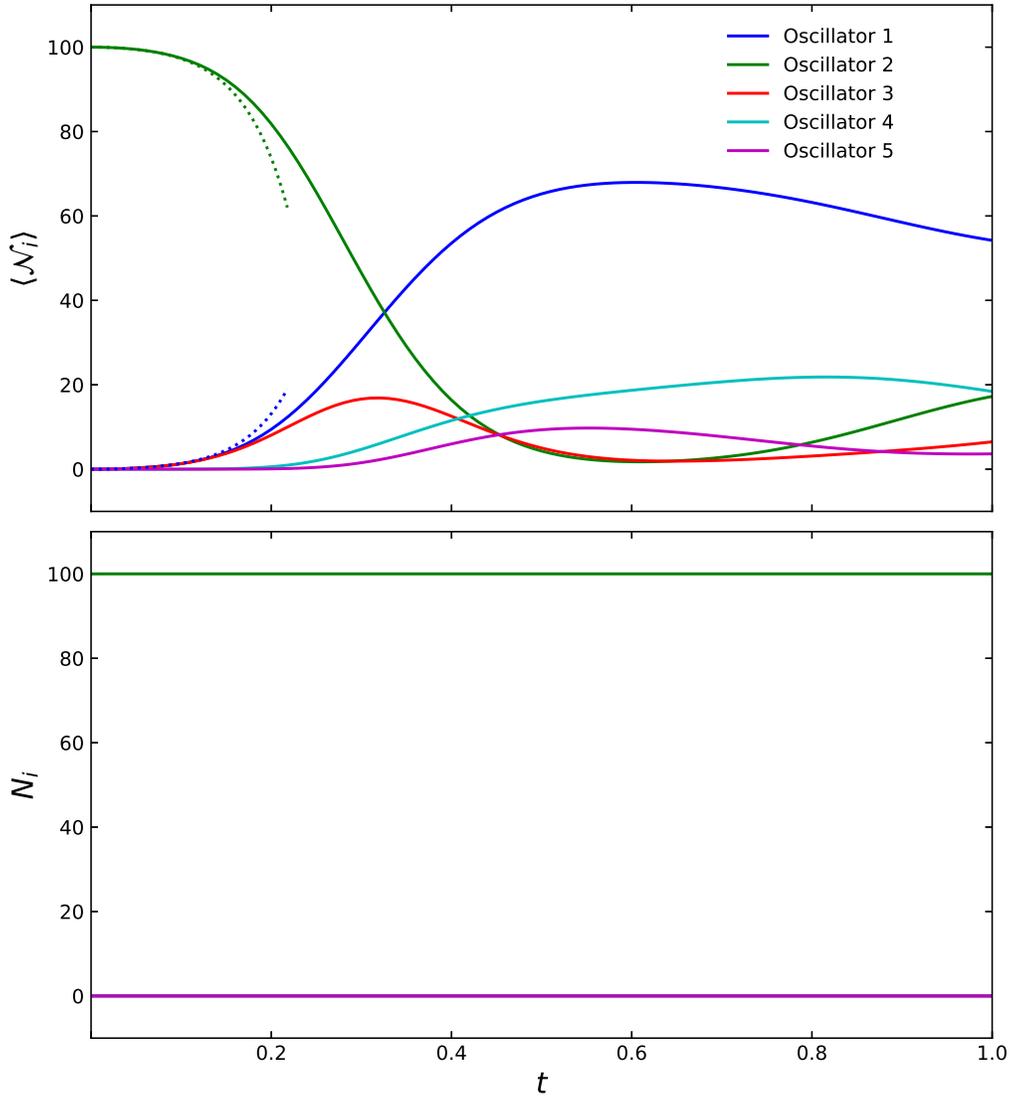}
\caption{Quantum (top) and classical (bottom) evolution 
of the initial state $|0, 100, 0, 0, 0 \rangle$.  In its
classical evolution this state persists indefinitely.  In 
its quantum evolution, the state thermalizes.  The dotted
lines show the predictions of Eqs.~(\ref{toy}) and 
(\ref{N2}).  After a time of order 0.1, these equations
are inaccurate because quanta jump from the 1st and 3rd 
oscillators back to the 2nd oscillator and from the 3rd 
to the 4th and 5th oscillators.}
\end{center}
\label{condensate}
\end{figure} 

\newpage

\begin{figure}
\begin{center}
\includegraphics[angle=360, height=50mm]{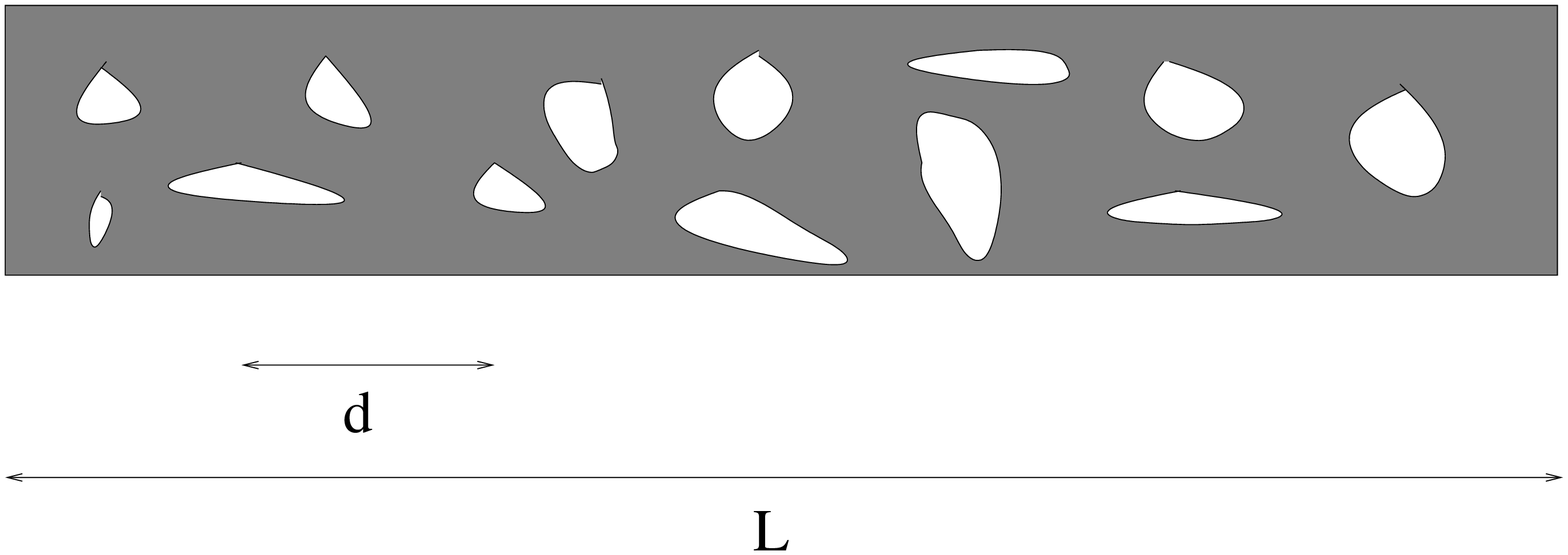}
\vspace{0.3in}
\caption{Cartoon of superfluid $^4$He filling a tube of 
length $L$ with various obstructions inside that make 
the fluid inhomogeneous on the length scale $d$. Although 
inhomogeneous on scale $d$ the fluid is correlated on scale 
$L$, which may be arbitrarily large compared to $d$.  Likewise 
the Bose-Einstein condensate of dark matter axions may be 
correlated on the scale of the horizon although inhomogeneous 
on the scale of galaxies.}
\end{center}
\label{cartoon}
\end{figure}



\begin{thebibliography}{bib}

\bibitem{PDM}
Reviews include: {\it Particle Dark Matter} edited by   
Gianfranco Bertone, Cambridge University Press 2010;
E.W. Kolb and M. Turner, {\it The Early Universe},
Addison Wesley 1990.

\bibitem{axion}
R. D. Peccei and H. Quinn, Phys. Rev. Lett. {\bf 38} (1977) 1440 and Phys.
Rev. {\bf D16} (1977) 1791; S. Weinberg, Phys. Rev. Lett. {\bf 40}
(1978) 223; F. Wilczek, Phys. Rev. Lett. {\bf 40} (1978) 279.

\bibitem{invax}
J. Kim, Phys. Rev. Lett. {\bf 43} (1979) 103; M. A. Shifman,
A. I. Vainshtein and V. I. Zakharov, Nucl. Phys. {\bf B166} (1980) 493;
A. P. Zhitnitskii, Sov. J. Nucl. {\bf 31} (1980) 260;  M. Dine,   
W. Fischler and M. Srednicki, Phys. Lett. {\bf B104} (1981) 199.

\bibitem{adm}
J. Preskill, M. Wise and F. Wilczek, Phys. Lett. {\bf B120} (1983) 127;
L. Abbott and P. Sikivie, Phys. Lett. {\bf B120} (1983) 133;
M. Dine and W. Fischler, Phys. Lett. {\bf B120} (1983) 137.

\bibitem{ipser}
J. Ipser and P. Sikivie, Phys. Rev. Lett. 50 (1983) 925.

\bibitem{Arias}
P. Arias et al., JCAP 1206 (2012) 013.

\bibitem{ULALP}
S.-J. Sin, Phys. Rev. D50 (1994) 3650;
J. Goodman, New Astronomy Reviews 5 (2000) 103;
W. Hu, R. Barkana and A. Gruzinov, Phys. Rev. Lett. 85 (2000) 1158;
E.W. Mielke and J.A. V\'elez P\'erez, Phys. Lett. B671 (2009) 174;
V. Lora et al.,JCAP 2 (2012) 011;
D.J.E. Marsh and J. Silk, MNRAS 437 (2014) 2652;
H.Y. Schive, T. Chiueh and T. Broadhurst, Nature Phys. 10 (2014) 496;
B. Li, T. Rindler-Daller and P. Shapiro, Phys. Rev. D89 (2014) 083536;
L. Hui, J.P. Ostriker, S. Tremaine and E. Witten, Phys. Rev. D95 (2017)
043541, and references therein.

\bibitem{CABEC}
P. Sikivie and Q. Yang, Phys. Rev. Lett. 103 (2009) 111301.

\bibitem{Christ}
N. Banik, A.J. Christopherson, P. Sikivie and 
E.M. Todarello, Phys. Rev. D95 (2017) 043542

\bibitem{axtherm}
O. Erken, P. Sikivie, H. Tam and Q. Yang,
Phys. Rev. D85 (2012) 063520.

\bibitem{Saik}
K. Saikawa and M. Yamaguchi, Phys. Rev. D87 (2013) 085010.

\bibitem{Berges}
J. Berges and J. Jaeckel, Phys. Rev. D91 (2015) 025020.

\bibitem{Khlopov}
M.Y. Khlopov, B.A. Malomed and Y.B. Zeldovich,
MNRAS 215 (1985) 575.

\bibitem{Bianchi}
M. Bianchi, D. Grasso and R. Ruffini, Astron. Astrophys. 231
(1990) 301.

\bibitem{inner}
A. Natarajan and P. Sikivie, Phys. Rev. D73 (2006) 023510.

\bibitem{crdm}
P. Sikivie, Phys. Lett. B432 (1998) 139.

\bibitem{sing}
P. Sikivie, Phys. Rev. D60 (1999) 063501.

\bibitem{MWhalo}
L. Duffy and P. Sikivie, Phys. Rev. D78 (2008) 063508.

\bibitem{case}
P. Sikivie, Phys. Lett. B695 (2011) 22.

\bibitem{Banik} 
N. Banik and P. Sikivie, Phys. Rev. D88 (2013) 123517.

\bibitem{Burkert}
A.M. Burkert and E. D'Onghia, Astrophys. and Space
Science Library 319 (2004) 341, and references therein.

\bibitem{Davidson}
S. Davidson and M. Elmer, JCAP 1312 (2013) 034.

\bibitem{Davidson2}
S. Davidson, Astropart. Phys. 65 (2015) 101.

\bibitem{Guth}
A.H. Guth, M.P. Hertzberg and C. Prescod-Weinstein,
Phys. Rev. D92 (2015) 103513.

\bibitem{simtherm}
P. Sikivie and E.M. Todarello, Phys. Lett. B770 (2017) 331.

\bibitem{Berges2}
J. Berges, K. Boguslavski, A. Chatrchyan and J. Jaeckel, 
arXiv:1707.07696.

\bibitem{Dvali}
G. Dvali and S. Zell, arXiv: 1710.00835.

\bibitem{Dvali13}
G. Dvali et al., Phys. Rev. D88 (2013) 124041.

\bibitem{Semikoz}
D.V. Semikoz and I.I Tkachev, Phys. Rev. D55 (1997) 489.

\bibitem{Bog}
N.N. Bogoliubov, J. Phys. (USSR) 11 (1947) 23, reprinted in 
D. Pines, {\it The Many-Body Problem}, W.A. Benjamin, New York
1961, p. 292.

\bibitem{Chav}
P.-H. Chavanis, A\&A 537 (2012) A127.

\end{thebibliography}
\end{document}